\documentclass[a4paper,11pt]{article}
\usepackage{jinstpub} 
\usepackage{lineno}
\usepackage[dvipsnames]{xcolor}
\usepackage{booktabs} 
\usepackage{placeins}
\usepackage{subcaption}
\usepackage{tikz}

\title{Extended thermal cycling of ATLAS ITk strip modules with and without stress mitigating interposers }

\author[a]{Nikolai Fomin,}
\author[a]{Bart Hommels,}
\author[a]{Thomas Ivison,}
\author[a]{Kosala Kariyapperuma,}
\author[a,b]{Jesse Liu}
 \affiliation[a]{Cavendish Laboratory, University of Cambridge, Cambridge CB3 0HE, UK}
\affiliation[b]{Department of Physics, New York University, New York City, NY 10003, USA}

\emailAdd{jesse.liu2@nyu.edu, hommels@hep.phy.cam.ac.uk}

\abstract{
This paper investigates critical mechanical failures during stand-alone thermocycling of ATLAS Inner Tracker strip pre-production modules. 
Five modules undergo extended thermocycling after adequately levelling thermal chucks and introducing interlocks for unattended operation. 
Module bow evolution is tracked via regular sensor metrology. 
All five modules exhibit bow increases with a mean of $146\pm 27~\mu$m  when raising maximum cycling temperatures from $20^\circ$C to $40^\circ$C. 
Four such modules exhibit sensor fractures when cycled to $-44^\circ$C. 
A stress-mitigating layer of silicone gel and Kapton film interposer is introduced to three further modules, with detailed quality control data establishing electromechanical viability. 
No significant bow change of $1\pm 10~\mu$m is observed after ten cycles between $[+40, -44]^\circ$C relative to $[+20, -44]^\circ$C.
Two interposer modules undergo 200 thermocycles up to $[+56, -44]^\circ$C without fracturing. 
This decreased sensor deformation and fracturing is interpreted as evidence for reduced thermal stress.
}


\begin{document}
\maketitle
\flushbottom

\clearpage
\section{\label{sec:intro}Introduction}
\FloatBarrier

The successful upgrade of the ATLAS Experiment~\cite{PERF-2007-01,ATLAS:2023dns} for the High-Luminosity Large Hadron Collider~\cite{Aberle:2749422} underpins future discoveries in particle physics~\cite{EuropeanStrategyGroup:2020pow,P5:2023wyd,Narain:2022qud,ATLAS:2019mfr,EuropeanStrategy:2019qin,ATL-PHYS-PUB-2018-031,ATL-PHYS-PUB-2018-005,ATL-PHYS-PUB-2021-039,ATL-PHYS-PUB-2024-001,ATL-PHYS-PUB-2025-006,Beresford:2019gww,Beresford:2024dsc}. 
Its cornerstone is a new all-silicon Inner Tracker (ITk) detector~\cite{ATLAS-TDR-PhaseII,ATLAS-TDR-30,ATLAS-TDR-25,ATLAS:2024rnw}. 
The ITk comprises pixel~\cite{ATLAS-TDR-30} and strip~\cite{ATLAS-TDR-25} sub-detectors, each designed as a barrel bookended by endcaps.
The strips barrel requires $10\ 976$ detector units called \emph{modules}, each comprising $97 \times 97$~mm p-type silicon strip sensors fabricated by Hamamatsu Photonics K.K.~\cite{Hommels:2016rli}.
Industrial adhesives~\cite{Poley:2015zza,Helling:2019zea,true-blue} mechanically attach sensors to custom electronics supplying front-end power, readout, monitoring, and control~\cite{Dabrowski:2009zz,Kaplon:2012wx,Michelis:2012uin,Diez:2014mra,Kuehn:2017cqo,Cormier:2021oog}.
Module assembly and testing within operating expectations, called quality control (QC)~\cite{ATLAS:2020ize,Tishelman-Charny:2024yys}, are distributed worldwide.

An important production QC procedure is \emph{module thermal cycling} (MTC).
Modules must pass electrical tests during repeated cycles between two temperatures.
The nominal operating window is between room ($+20^\circ$C) and end-of-life temperatures ($-35^\circ$C) to mitigate elevated leakage currents from radiation damage~\cite{ATLAS:2021zxb,ATLAS:2021gld,Dawson:2019cmh}.
Every module undergoes ten thermocycles to verify electrical functionality and mechanical integrity between $[+20, -35]^\circ$C. 
Cooling or power failures~\cite{Arling:2022dru} can expose modules to temperatures outside the operating window. 
This motivates studies beyond nominal QC temperatures. 
Testing outside operating expectations is called quality assurance (QA).

\begin{figure}
    \centering
    \begin{subfigure}{0.2\textwidth}
    \centering
    \includegraphics[height=3.1cm]{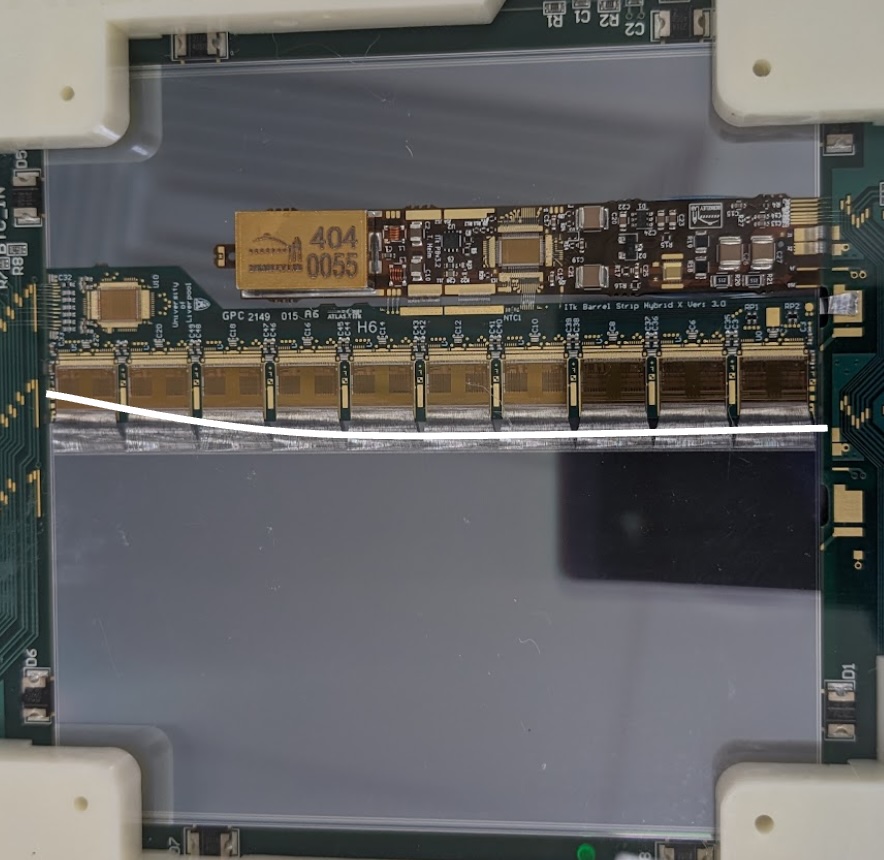}
    \caption{Fracture line}
    \end{subfigure}
    \begin{subfigure}{0.79\textwidth}
    \centering
    \includegraphics[height=3.1cm]{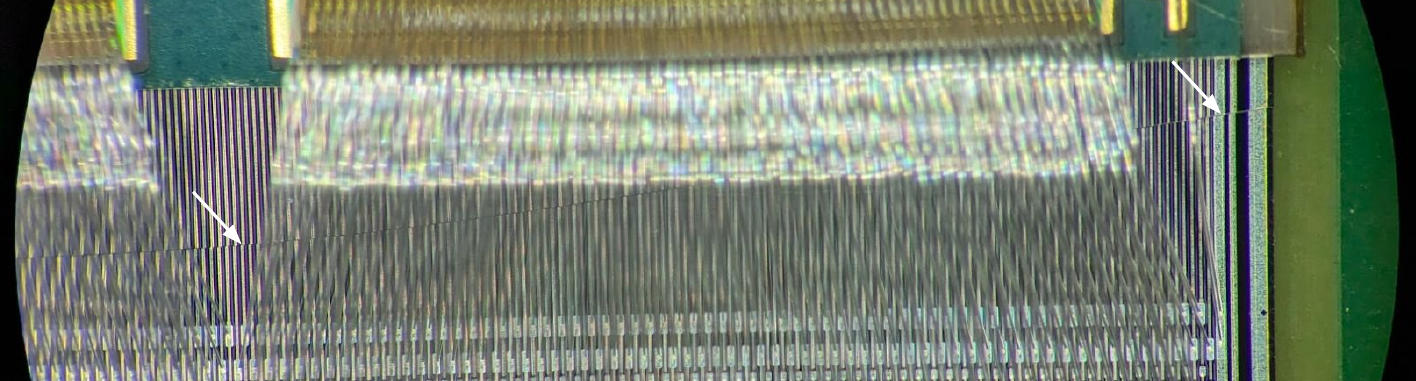}
    \caption{Zoom into right edge with arrows showing fracture}
    \end{subfigure}
    \caption{Photo of module 20USBML1234911 thermocycled between $[+40, -35]^\circ$C prior to this paper, showing a sensor fracture. 
    (a) The white line indicates the fracture position on the sensor. 
    (b) Zoom into the right edge has arrows indicating the fracture under the wire bonds.}
    \label{fig:ppb2-ls06-crack}
\end{figure}

However, we observe critical problems when thermocycling ten pre-production modules between $+40^\circ$C and $-35^\circ$C: visual inspection identifies a sensor fracture for one module (figure~\ref{fig:ppb2-ls06-crack}) and three modules develop excess leakage currents during $-35^\circ$C tests.
Independent efforts report similar failure modes but only \emph{after} loading modules onto local support structures, which is called the \emph{sensor fracturing} problem~\cite{DiezCornell:2899168,Tishelman-Charny:2024wxu}.
Simulation studies~\cite{itk-strips-cracks-simulation} suggest adverse stress from mismatched coefficients of thermal expansion (CTE)~\cite{pcb-cadence,DELUCA2011549,Danilewsky:xz5004,Vellukunnel:10100842} between the silicon sensor (fixed to local supports with adhesives) and copper in the flexible printed circuit board (flex PCB); the intermediate epoxy glue provides a high-modulus mechanical coupling that transfers stress.

\begin{figure}
    \centering
    \begin{subfigure}{0.75\textwidth}
    \centering
    \includegraphics[width=\textwidth]{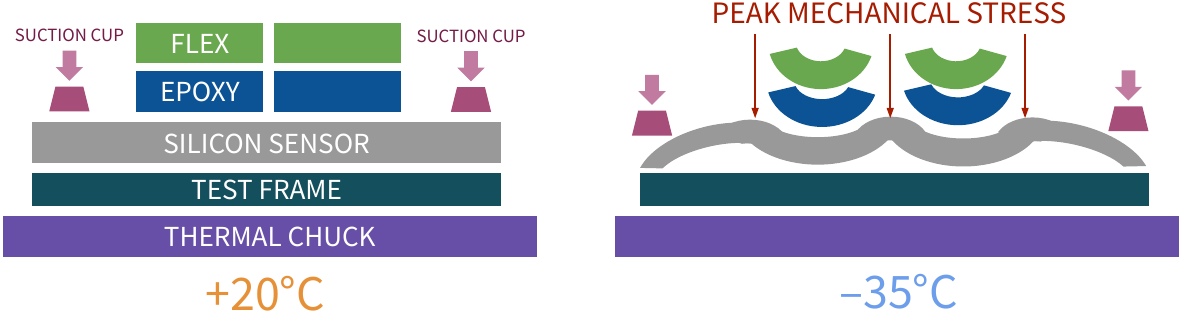}
    \caption{Stackup of pre-production module installed in thermocycler}
    \end{subfigure}
    \begin{subfigure}{0.24\textwidth}
    \centering
    \includegraphics[width=\textwidth]{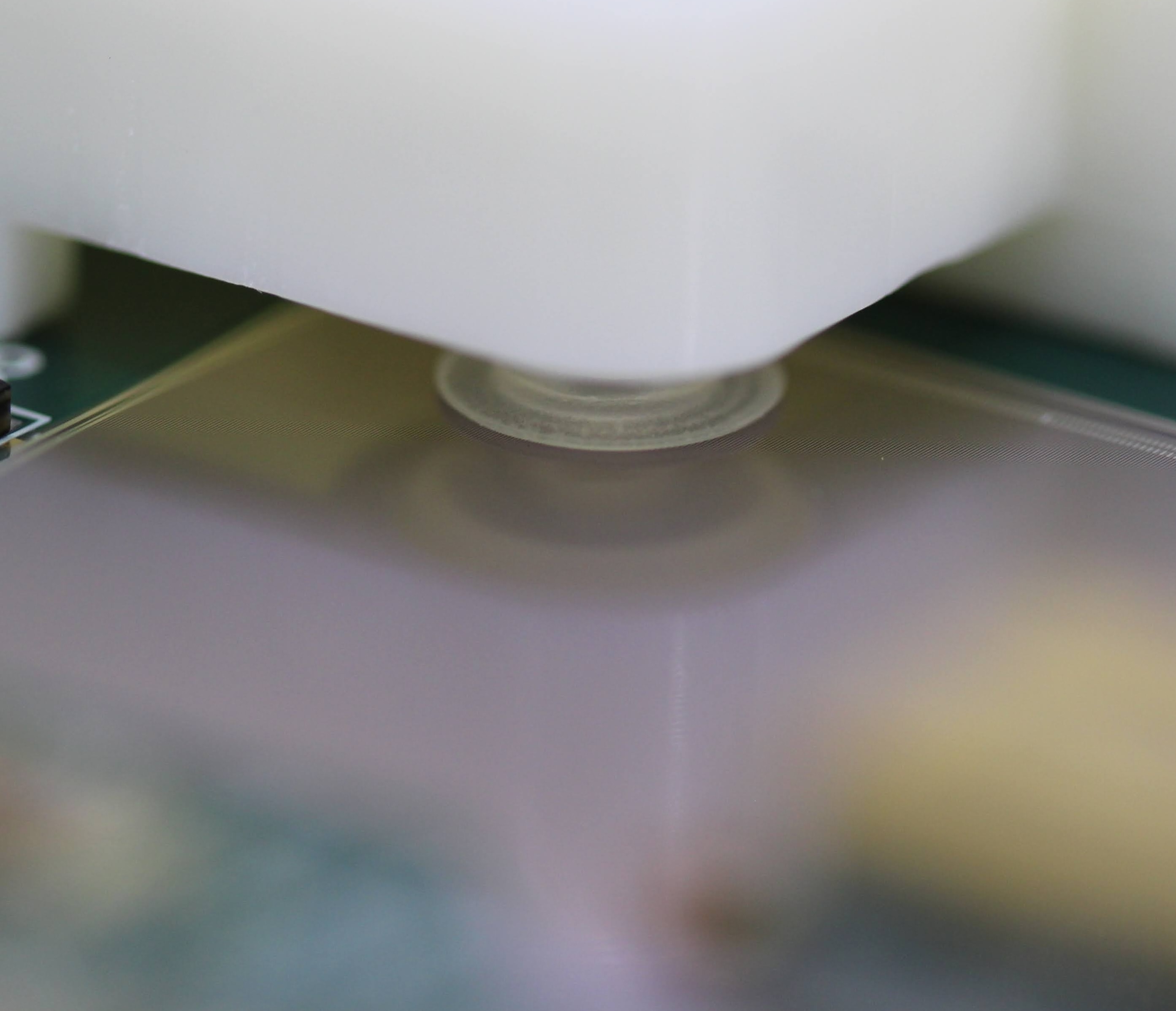}
    \caption{Suction cup on sensor}
    \end{subfigure}
    \caption{\label{fig:module-ppb-stackup} (a) Schematic showing the stackup of a pre-production module  installed in the module thermocycling apparatus at $+20^\circ$C (left) and $-35^\circ$C (right) chuck temperatures. 
    Mismatched thermal expansion coefficients cause the flex PCB, epoxy adhesive, and sensor to contract unevenly.
    This induces mechanical stress at $-35^\circ$C localised at the epoxy-sensor boundaries. 
    This adapts the diagrams in Ref.~\cite{itk-strips-cracks-simulation} to module thermocycling, where a handling frame (not shown) has suction cups fixing the sensor to the test frame rather than adhesives onto local supports. (b) Close-up photo of a suction cup contacting the sensor surface. }
\end{figure}

These critical failures preclude production. 
Thermal stress is hypothesised to be linked to prior MTC fracturing (figure~\ref{fig:ppb2-ls06-crack}), where suction cups induce a large lever arm on the deformed sensor of cooled modules (figure~\ref{fig:module-ppb-stackup}). 
Site-specific apparatus differences may also indicate why we observe sensor fracturing during MTC, while other sites only report systematic fracturing after loading modules onto local supports. 
It is important to test these hypotheses and resolve site-specific MTC failures as a separate problem from post-MTC fracturing in the wider community. 

This paper presents in-house investigations and mitigations for these critical problems.
Section~\ref{sec:setup} briefly reviews module assembly and the setup. 
Section~\ref{sec:Cam-developments} introduces custom hardware interlocks, enabling unattended operation and extended thermal cycling studies. 
While investigating the prior sensor fracture (figure~\ref{fig:ppb2-ls06-crack}), we find non-flat chucks impart extraneous mechanical stress on modules during MTC and propose custom shims to level the chucks.
Section~\ref{sec:non-interposer-thermocycling} uses this improved setup for a test-to-destruction QA study, similar to Ref.~\cite{Tishelman-Charny:2024jnc}, to systematically reproduce the conditions for mechanical failure in figure~\ref{fig:ppb2-ls06-crack}. 
Permanent sensor deformations are observed after MTC when raising the chuck above $30^\circ$C~\cite{Salami:2025nob}.
This hysteresis is hypothesised to arise from a glass transition in the epoxy~\cite{Carbas02012014} that ``bakes in'' deformations induced by stress from the CTE mismatch.
We exploit this effect by using the sensor bow of free modules as the main observable in this study. 

Section~\ref{sec:interposers} introduces stress-mitigating materials onto modules to test the thermal stress hypothesis causing sensor deformations during MTC. 
Simulation studies~\cite{itk-strips-cracks-simulation} predict reduced thermal stress by inserting two materials between the flex and epoxy called \emph{interposers}: 
(i) 100~$\mu$m thick layer of SE-4445 silicone gel~\cite{dowsil-se-4445} whose low modulus decouples stress~\cite{Matayabas:2005,larson-dow,Fortman:2926236} induced by CTE mismatch between the flex PCB and sensor, and
(ii) 50~$\mu$m thick layer of Kapton polyimide film~\cite{dupont-kapton} to ensure adhesion~\cite{Jeong:2021,YUN1997827} with the epoxy.
We propose an ad hoc procedure to attach interposers onto modules. 
To test its engineering feasibility, we subject these trial interposer modules to the same QC and extended QA thermocycling as pre-production modules.
This probes potential adverse side effects of these materials and ad hoc assembly, including the long-standing \emph{cold noise} problem~\cite{Dyckes:2891402}, while testing for reduced sensor deformation hypothesised from simulated stress mitigation. 

\section{\label{sec:setup}Module assembly and setup}

\begin{figure}
    \centering
    \begin{subfigure}{0.49\textwidth}
    \centering
    \includegraphics[height=5.5cm]{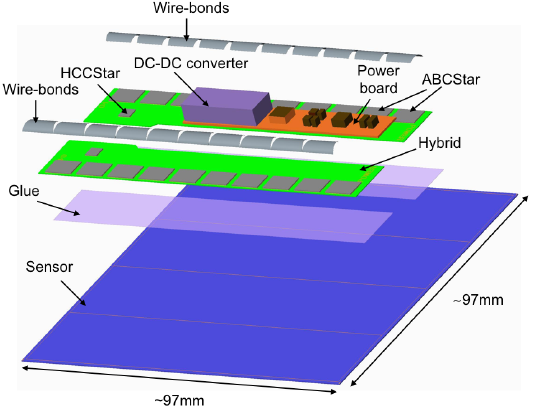}%
    \caption{\label{fig:pp-SS-module-design}Pre-production module schematic from Ref.~\cite{ATLAS-TDR-25}}
    \end{subfigure}
    \begin{subfigure}{0.49\textwidth}
    \centering
    \includegraphics[height=5.5cm]{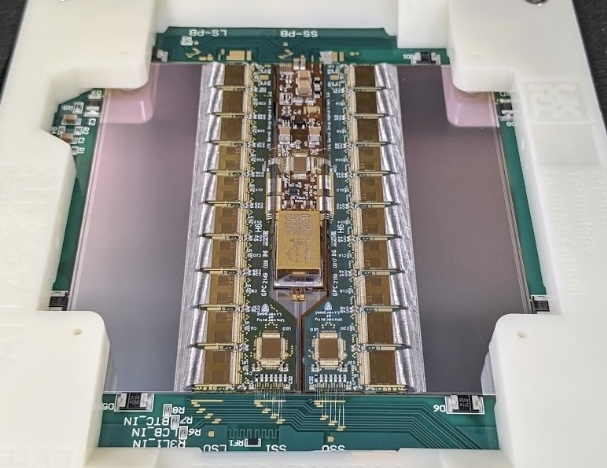}
    \caption{\label{fig:assembled-SS-module}Photo of module SS-04 with interposers}
    \end{subfigure}
    \caption{
    (a) Short-strip (SS) module without interposers showing custom microelectronics (HCCStar, ABCStar~\cite{Lu:2017jnv,ATLAS:2023sqb}) on flexes glued to the sensor.
    (b) Photo of assembled interposer short-strip module (SS-04) sitting on the green test frame with a white handling frame housing four suction cups holding the sensor down on each corner. 
    The amber Kapton film of the interposer is visible surrounding the flex edges. 
    Long-strip (LS) modules have the same features but one fewer hybrid.
    }
\end{figure}

\begin{figure}
    \centering
    \includegraphics[width=\textwidth]{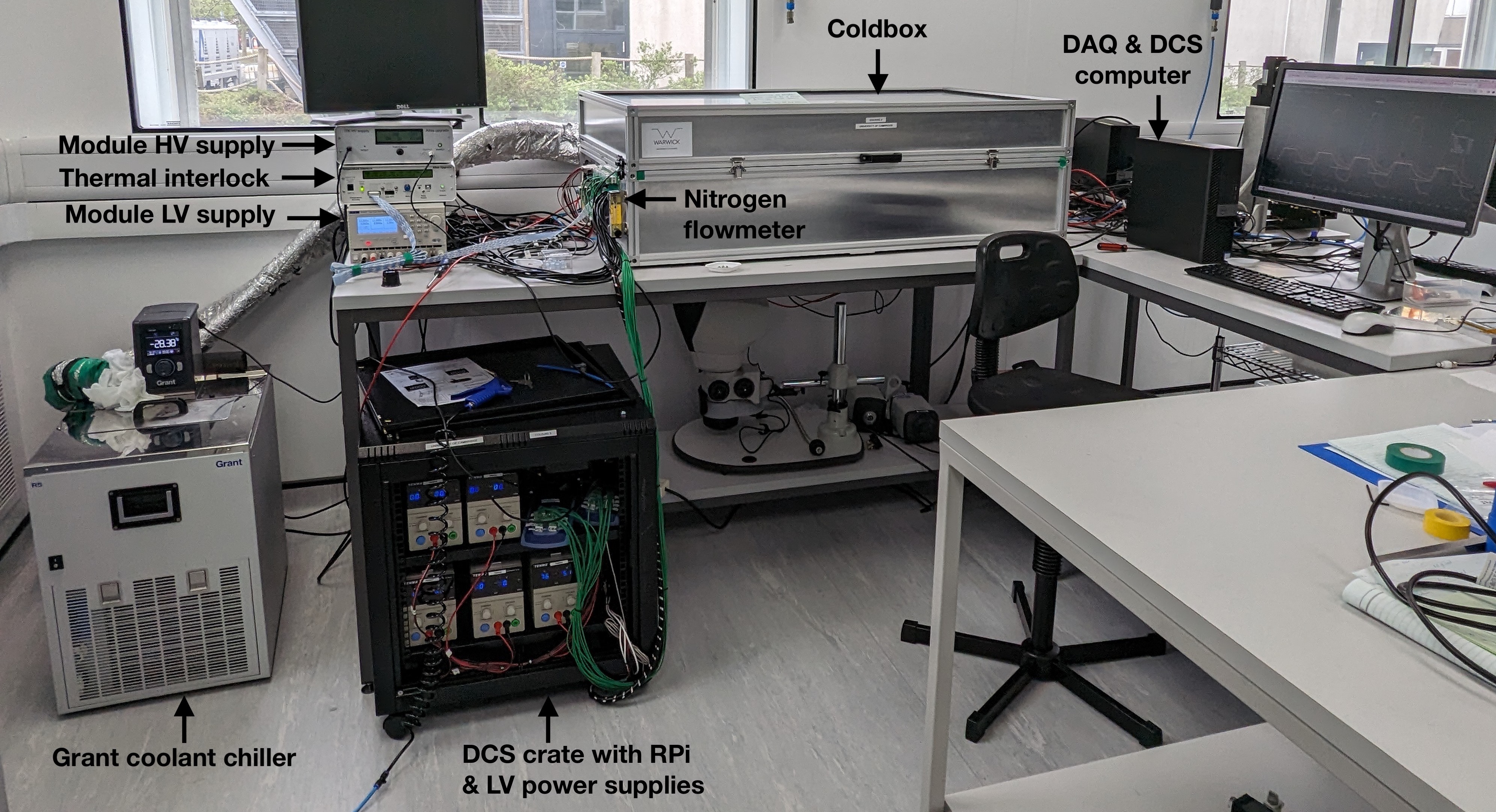}
    \caption{\label{fig:MTC_setup}
    Photo of module thermocycling setup.
    Section~\ref{sec:Cam-developments} describes the custom HV power supply and thermal interlock systems (figure~\ref{fig:itk-mtc-interlocks}).}
\end{figure}

We briefly review the pre-production module assembly procedure and thermocycling setup.
Figure~\ref{fig:pp-SS-module-design} 
illustrates the original pre-production barrel module design.
Pre-production assembly uses \textsc{Loctite Eccobond F112}~\cite{true-blue} epoxy resin\footnote{The ITk Strip community refers to this adhesive as \emph{TrueBlue} due to its optical colour.}.
This attaches the sensor to \emph{hybrid(s)} and \emph{powerboard} PCBs housing custom microelectronics~\cite{Lu:2017jnv,ATLAS:2023sqb,Villani:2020mqp,Gosart:2023pcl} for front-end readout, control, and power. 
There are two types of modules: \emph{long strip} (LS) that have one hybrid to read out 2560 channels, and \emph{short strip} (SS) that have two hybrids to read out 5120 channels. 
After wire bonding, modules undergo QC testing~\cite{ATLAS:2020ize} in the coldbox of the MTC setup situated at the University of Cambridge (figure~\ref{fig:MTC_setup}). 
This apparatus combines thermoelectric Peltiers with liquid coolant circulated by a \textsc{Grant} chiller~\cite{grant-chiller,grant-chiller-Rmanual,grant-chiller-txf200}, interfaced with custom detector control system (DCS) and data acquisition (DAQ)~\cite{pi-plate,xilinx-nexys}, with continuous environmental monitoring.

This paper probes mechanical deformations by measuring the sensor bow of free modules outside the MTC setup.
We employ the \textsc{Baty Venture 3030} coordinate-measuring machine (CMM)~\cite{baty-venture-3030}. 
We develop an in-house script to convert the machine output into a Common Data Format (CDF) defined by standard QC procedures~\cite{ATLAS:2020ize,Salami:2025nob}. 
A community-developed script then reads this CDF file and computes the sensor bow using a standard bow algorithm. 
The algorithm first performs a fit to all sensor points $p_i = (x_i, y_i, z_i)$ to define a sensor plane $\Pi$.
The coordinate system convention is defined in the caption of figure~\ref{fig:cracks-guides}. 
To remove any global sensor tilt, the residuals $z'_i = z_i - z_\Pi$ relative to the sensor plane are used to calculate the bow magnitude $|z_\text{bow}|$ as the range of residuals:
\begin{equation}
  |z_\text{bow}| = \text{max}(z_i') - \text{min}(z_i').
  \label{eq:bow-definition}
\end{equation}
The sensor is divided into $N$ sections in $x$ and $y$ to impose an $N\times N$ grid over sensor points.
The algorithm then defines $p_\text{edge}$ as the subset of points in the outer ring of the grid while $p_\text{centre}$ comprises all other points.
The sign convention is positive if the mean height of the centre points $\bar{z}_\text{centre}$ is greater than the mean height of the edge points $\bar{z}_\text{edge}$, else it is negative.

\section{\label{sec:Cam-developments}Improvements to module thermocycling apparatus}

To ensure modules stay within pre-defined operating conditions during MTC, we develop in-house stand-alone interlocks.
These supplement the existing system originally designed and constructed by the University of Warwick.
This enables safe, unattended and overnight operation crucial for realising the extended thermocycling studies proposed in this paper.

\begin{figure}
    \centering
    \includegraphics[width=\linewidth]{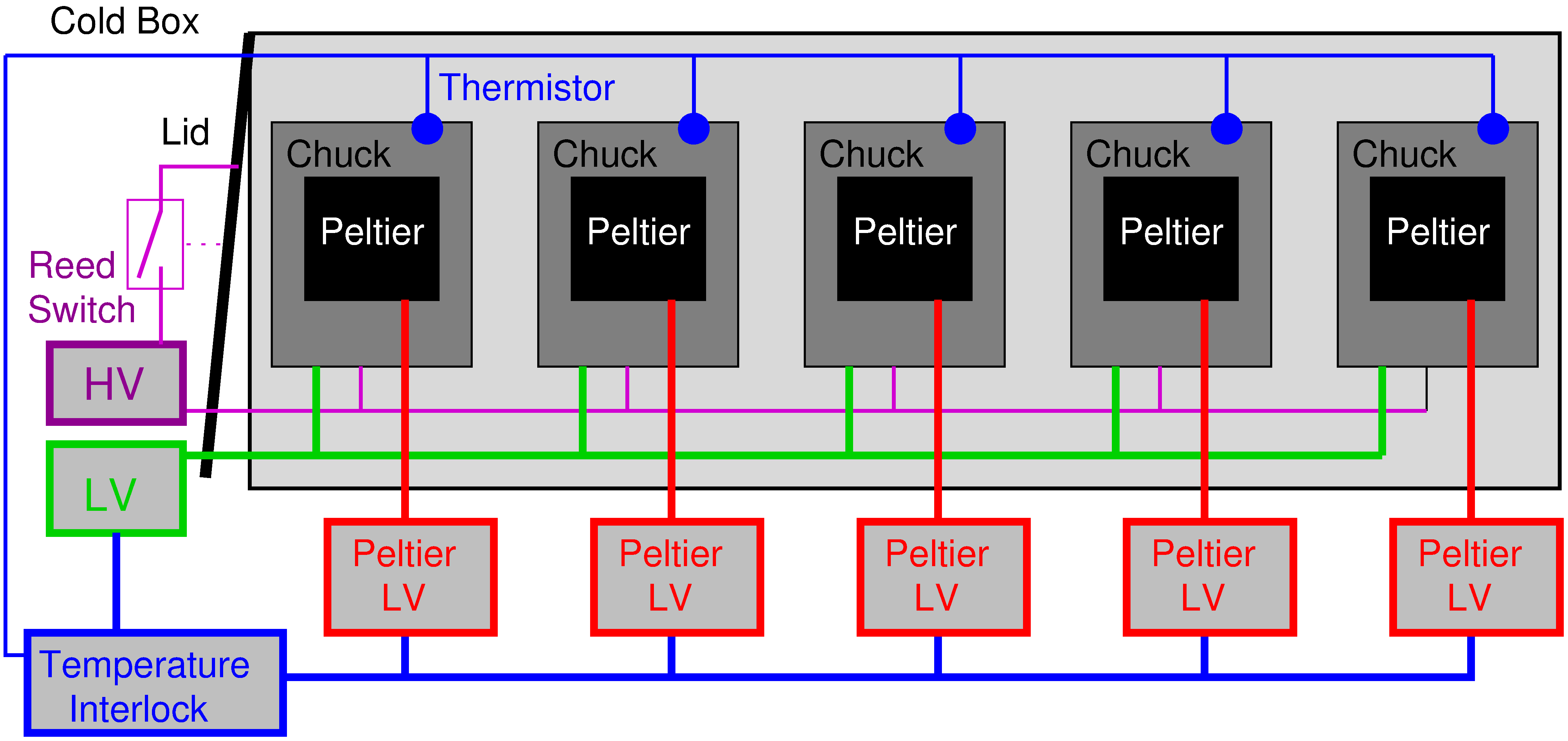}
    \caption{Schematic of the MTC coldbox, where Peltier LV supplies (red) power thermoelectric plates attached to each chuck (grey).
    We introduce in-house hardware: 
    (i) the high-voltage (HV) power supply (purple) provides bias voltage for sensors, whose interlock is triggered by opening the lid connected to a reed switch, 
    and (ii) the thermal interlock (blue) cuts power to the module (green) and Peltier LV supplies when thermistors attached to chucks (figure~\ref{fig:modules-ppc-ls}) sense overheating events. }
    \label{fig:itk-mtc-interlocks}
\end{figure}

\subsection{\label{subsec:Tinterlock}Thermal interlock}
To protect both the modules and MTC setup against temperature excursions outside the safe range, we install a stand-alone thermal interlock system~\cite{temperature-interlock}.
It comprises custom electronics to read negative temperature coefficient (NTC) thermistors attached to the thermal chucks, an \textsc{Arduino Uno} microprocessor board, and switches to turn on and off two mains feed-through channels. 
The interlock cuts the mains from the Peltier power supplies when any thermistor reading exceeds the set range (figure~\ref{fig:itk-mtc-interlocks}). 
This protects against situations where the thermocycler DCS is in an unknown state while the Peltier elements are powered, leading to uncontrolled temperatures. 

To avoid self-heating beyond their operating limits from module power dissipation, the interlock cuts out the low-voltage module power supply. 
The current rating of the built-in mains switches is insufficient to safely switch the power to the main chiller unit. 
This is deemed unnecessary since the chiller thermal range is intrinsically limited; additionally, it has two levels of built-in interlocks to prevent damage to the coldbox or the chiller itself.

\subsection{\label{subsec:ITkHV}High-voltage supply and interlock}

We also develop a custom HV power supply as an add on to the MTC setup.
It is based on a \textsc{Spellman} MPS DC-to-HV converter capable of delivering 2.5~mA at $-700$~V.
It uses an \textsc{Arduino} Leonardo microprocessor to operate the display, facilitate over-current and safety interlock functionality, and provide a serial-over-USB interface for remote readout and control.
The HV converter has monitoring outputs for voltage and current, which are accurate to 1\% of the full scale only. 
The unit features a dedicated interlock by interfacing to a magnetic reed switch attached to the coldbox lid (figure~\ref{fig:itk-mtc-interlocks}).
When the lid is opened with HV enabled, the interlock rapidly reduces the HV output to 0~V at 100~V~s$^{-1}$. 
This introduces added safety beyond the existing software interlock.

\subsection{\label{subsec:Tchuck}Improved thermal chuck flatness}

\begin{figure}
    \centering
    \begin{subfigure}{0.49\textwidth}
    \centering
        \includegraphics[trim={0.1cm 0.1cm 0.1cm 0.1cm},clip,width=\textwidth]{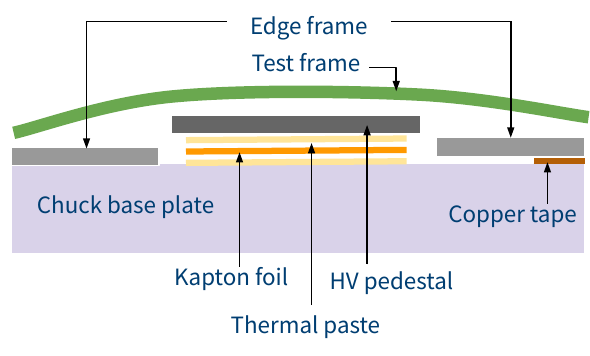}
        \caption{Original thermal chuck stackup}
    \end{subfigure}
    \begin{subfigure}{0.49\textwidth}
    \centering
        \includegraphics[trim={0.1cm 0.1cm 0.1cm 0.1cm},clip,width=0.99\textwidth]{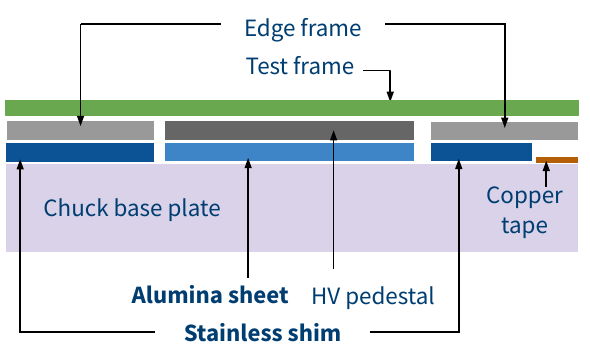}
        \caption{In-house modifications to level chuck}
    \end{subfigure}
    \caption{(a) Schematic of chuck pedestal stackup in the original module thermocycling apparatus. 
    (b) In-house designed alumina sheet and stainless shim (blue) to level thermal chuck.  }
    \label{fig:mtc-chuck-modification}
\end{figure}

Motivated by prior sensor fracturing (figure~\ref{fig:ppb2-ls06-crack}), we investigate potential apparatus problems.
We identify the HV pedestal relative to edge frame of the thermal chucks to be the greatest source of non-flatness.
This was overlooked as chuck flatness was not in the original hardware specification. 
Figure~\ref{fig:mtc-chuck-modification} shows a schematic of the thermal chucks with module loaded and the central pedestal being higher than the side. 
Table~\ref{tab:chuck-heights} displays measurements using feeler gauges for the different chucks in our MTC hardware.
The mean level difference is $200 \pm 60~\mu$m with a maximum of 300~$\mu$m, which exacerbates the stress imparted on modules installed on the chuck. 

To remedy this, we propose and design an aluminium oxide (alumina) sheet to sit underneath the central HV pedestal, together with a stainless shim to sit underneath the outer piece of the chuck. 
The custom stainless shim levels the edge frame while providing a low-impedance electrical contact to neighbouring chucks.
The alumina shim of equal thickness replaces the stack of thermal paste and Kapton film causing the lack of flatness.
It is desirable for the pedestal to sit higher than the frame by 100 to 200~$\mu$m to ensure a good thermal and electrical contact between the chuck, the module test frame PCB and the module itself.
We fit shims designed in house  to improve flatness and target this residual level difference.
This avoids air gaps deteriorating the module-chuck thermal coupling and build-up of potential differences leading to electrical discharges or sparks.

After fitting the shims and electrical connection pieces between the shims, feeler gauges determine the residual height differences with results in table~\ref{tab:chuck-heights}.
The post-shim values are all lower relative to the corresponding point before shimming, where the mean reduces to $60\pm 50~\mu$m with a maximum 170~$\mu$m. 
This demonstrates improved chuck flatness relative to the original design. 
A bow of up to $100~\mu$m is deemed tolerable for subsequent use. 
Comparative metrology measurements indicate a high degree of similarity between the bow of a module resting on a flat gauge plate and a module resting on a test frame bolted to a known flat jig surface. 
For consistency in the upcoming studies, sensor bow metrology is performed with the module resting on the same test frame at room temperature.

\begin{table}[]
\centering
\includegraphics[height=2.7cm]{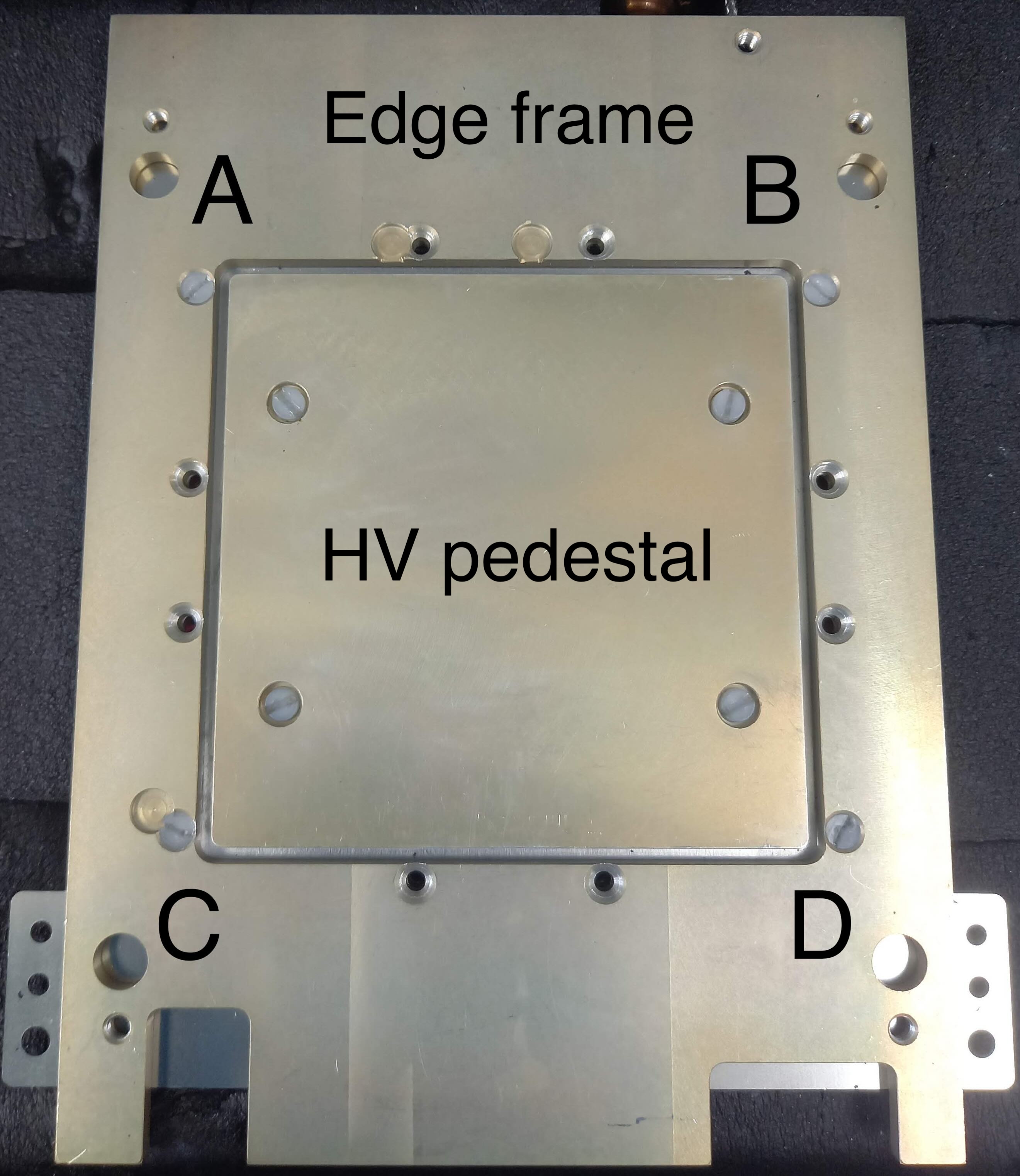}%
\quad
\begin{tabular}{@{}ccccc|cccc@{}}
\toprule
      &         \multicolumn{4}{c}{Before shim $\Delta z_{\text{centre}, \text{edge}}~[\mu\text{m}]$ } & \multicolumn{4}{c}{After shim $\Delta z_{\text{centre}, \text{edge}}~[\mu\text{m}]$ } \\ 
Chuck & A    & B      & C      & D    & A   & B   & C   & D             \\ \midrule
1     & 200   & 300    & 100   & 200  & 80  & 0   & 20  & 50              \\
2     & ---   & ---    & ---   & ---  & 0   & 0   & 50  & 100             \\
3     & 200   & 200    & 100   & 200  & 170 & 0   & 100 & 120             \\
4     & 250   & 250    & 250   & 250  & 160 & 10  & 10  & 10             \\
5     & 150   & 150    & 200   & 250  & 20  & 100 & 40  & 70             \\\bottomrule
\end{tabular}
\caption{\label{tab:chuck-heights}Residual heights between the chuck central pedestal and edge in our MTC apparatus before and after shimming (figure~\ref{fig:mtc-chuck-modification}). 
No data are measured for chuck 2 before shimming.
}
\end{table}

\section{\label{sec:non-interposer-thermocycling}Module thermocycling to destruction}

This section introduces the module QA study to probe the unexpected failure modes of figure~\ref{fig:ppb2-ls06-crack}.
This also verifies the improved MTC setup lacks adverse side effects and alleviates apparatus-induced stress. 
A leading hypothesis for module thermal stress is the CTE mismatch of the flex-epoxy-sensor components. 
While numerical simulation can guide diagnosis and mitigation~\cite{itk-strips-cracks-simulation}, laboratory data are indispensable for empirical validation. 
To test these hypotheses, we propose measuring the sensor bow vs cycling temperature in our MTC setup.
We systematically increase the module thermal stress by progressively widening the thermocycling temperature range.

This QA study has two principal objectives. 
First, we reproduce the conditions for sensor fracturing in figure~\ref{fig:ppb2-ls06-crack} in a controlled procedure by systematically widening the MTC temperature range. 
Second, we measure sensor bow after every few thermocycles to monitor deformations evolving with extended thermocycling. 
This establishes baseline control data, wherein proposed solutions must demonstrate reduced sensor deformation under similar thermal conditions (section~\ref{sec:interposers}). 
It also informs why modules fail only after repeated thermal cycling instead of at the first cycle.

\subsection{\label{sec:design_mtc_study}Test-to-destruction methodology}

\begin{table}[]
\centering
\begin{tabular}{cccc}
\toprule
\multicolumn{2}{c}{No interposers} & \multicolumn{2}{c}{With interposers} \\
\midrule
Local name     & Serial number     & Local name      & Serial number      \\ \midrule
LS-01          & 20USBML1234912    & SS-04           & 20USBMS0000302     \\
LS-02          & 20USBML1235184    & SS-05           & 20USBMS0000300     \\
LS-03          & 20USBML1235196    & LS-12           & 20USBML1235389     \\
LS-04          & 20USBML1234747    &                 &                    \\
LS-06          & 20USBML1235267    &                 &                   \\ \bottomrule
\end{tabular}
\caption{\label{tab:summary_ppc_ls_modules} 
Summary of modules assembled for this paper by local name and module serial number.  
Five pre-production long-strip (LS) modules have no interposers (section~\ref{sec:non-interposer-thermocycling}). Three interposer modules are assembled comprising two short-strip (SS) and one LS (section~\ref{sec:interposers}).}
\end{table}

\begin{figure}
    \centering
    \includegraphics[width=\textwidth]{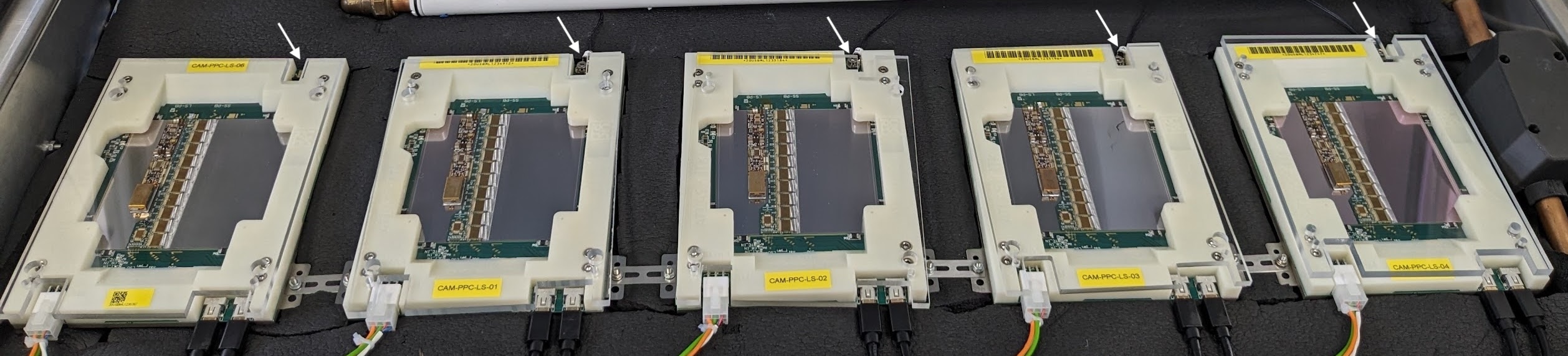}
    \caption{\label{fig:modules-ppc-ls}Photo of five pre-production long-strip modules mounted onto chucks for thermocycling after mitigating chuck flatness.
    The low and high voltage power is delivered on the bottom left of each module via a Molex connector. 
    DisplayPort connections provide data readout and control connections at the bottom right of each module.
    White arrows point to thermistors for the thermal interlock, which connect to a ribbon cable (not shown) attached to the underside of the leftmost chuck, and for the remaining chucks via an M4 screw to the top-right of the test frame. 
    }
\end{figure}

Table~\ref{tab:summary_ppc_ls_modules} summarises the modules assembled with and without interposers for this paper. 
Test-to-destruction goals and limited sensor availability motivate us to select ``B-grade'' sensors for non-interposer modules that formally fail sensor QC criteria~\cite{Miyagawa:2024stg}. 
Specifically, these sensors exhibit modestly elevated leakage currents $\gtrsim 500$~nA at 500~V bias voltage\footnote{This paper follows the convention of quoting the absolute magnitude of the bias voltage.}.
This loosened electrical criterion does not affect the sensor mechanical properties, enabling reliable bow studies without consuming valuable production components.
After verifying these pre-production modules are assembled to specification, they are installed onto the MTC chucks (figure~\ref{fig:modules-ppc-ls}). 
The standard MTC sequence is ten cycles between $[T_\text{max}, T_\text{min}] = [+20, -35]^\circ$C chuck temperatures at no more than $2.5^\circ$C per minute rate~\cite{Tishelman-Charny:2024yys}. 
To verify test conditions, 
continuous monitoring of environmental data is available for each module. 
A set of DAQ tests lasting around 10--15 minutes is taken at $T_\text{max}$ and $T_\text{min}$ with sensors biased if leakage currents do not trigger HV interlocks.

The proposed test-to-destruction QA study has the following methodology:
\begin{enumerate}
\item \textbf{Sensor bow of free module after thermocycling}. 
We measure the sensor bow of a free module \emph{ex situ} outside the coldbox in ambient cleanroom conditions using our CMM, which indicates deformations induced by MTC. 
We lack the instruments to perform \emph{in situ} metrology at cold or warm temperatures while the module is thermocycled.

\item 
\textbf{Baseline bow with [+20, $-$35]$^\circ$C thermocycling}. 
We establish baseline control data of sensor bow thermocycling after $N_\text{cycles} = 1, 5$, and 10 in the nominal QC range $[T_\text{max}, T_\text{min} ]= [+20, -35^\circ]$C. 
No bow change is expected in this range as maximum temperatures do not exceed that of the epoxy glass transition.

\item \textbf{Raise $T_\text{max}$ by 5$^\circ$C increments}. 
Increasing $T_\text{max}$ in $5^\circ$C increments, we expect the free-module bow to increase significantly from $T_\text{max} = 20^\circ$C to $T_\text{max} = 40^\circ$C.
This estimates the critical temperature $T_\text{crit}$ within the $5^\circ$C step resolution that triggers sensor deformation.

\item \textbf{After $T_\text{max}$ reaches 40$^\circ$C, decrease $T_\text{min}$ to $-$44$^\circ$C}. 
We determine $ -44^\circ$C to be the lowest stable temperature achievable for all five modules in our MTC setup.
Lower temperatures induce greater thermal stress in the module due to CTE mismatch. 
We keep module low-voltage enabled throughout MTC to produce realistic module thermal gradients and bias the sensors at $300$~V for modules without early HV breakdown. 

\item \textbf{Extend thermocycling until fractures form}.
Sensor deformation from thermal stress accumulates in the sensor bow with repeated cycles, which can lead to sensor fractures.
We evaluate electrical data and perform sensor visual inspection to search for fracture formation.
\end{enumerate}

\begin{figure}
    \centering
    \includegraphics[width=\textwidth]{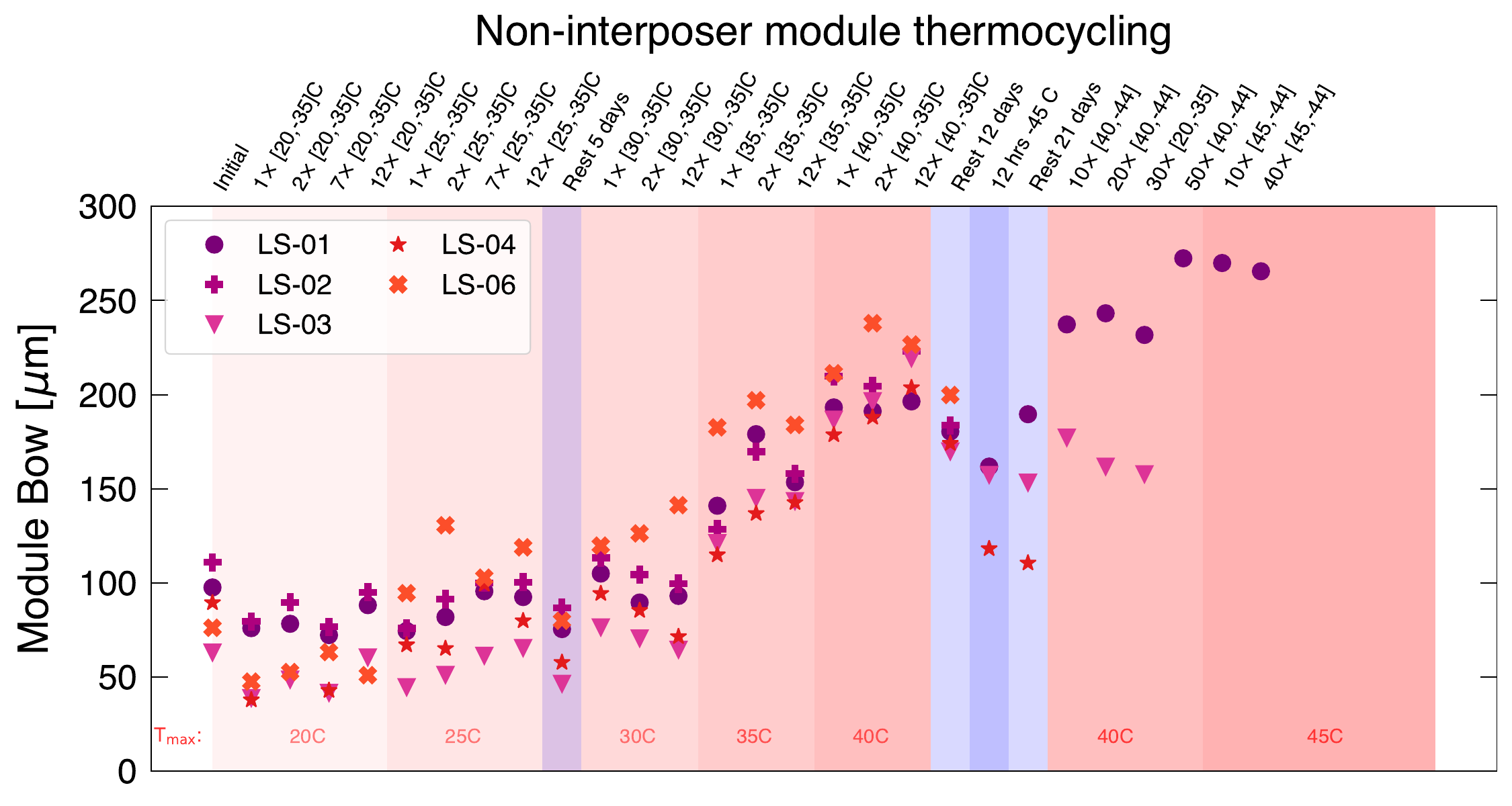}
    \caption{\label{fig:bowVsTemp-nointerposer} 
    Summary of sensor bow with thermocycling step for five pre-production modules (filled markers).
    Axis labels show $N\times$ accumulated cycles between $[T_\text{max}, T_\text{min}]$ temperatures.
    Each marker corresponds to a subsequent bow measurement. 
    Red shading shows the maximum temperature $T_\text{max}$ labelled on the plot. 
    Blue shading shows bow metrology after resting pre-production modules in dry storage for the number of days in the upper axis. 
    }
\end{figure}

\subsection{\label{subsec:bowvscycling}Bow evolution versus cycling temperature}
We now present sensor bow results with cycling temperatures for the five pre-production modules summarised in figure~\ref{fig:bowVsTemp-nointerposer}. Discussion of interposer modules results is deferred to section~\ref{subsec:evidence-mitigated-stress}.

We first report the regime with no sensor deformation. 
After the first cycle between $[+20, -35]^\circ$C, we find a minor bow decrease of around $30~\mu$m and the trends are correlated across all five modules, suggesting mechanical relaxation.  
For simplicity, we average the bow data over the five modules after $[+20, -35]^\circ$C MTC to yield $\bar{z}_\text{bow} = 68\pm 23~\mu$m (table~\ref{tab:table-bow-diff-mean}).
Thermocycling between $[+25, -35]^\circ$C induces little change in bow, except LS-06 rises anomalously to $131~\mu$m after the second $[+25, -35]^\circ$C cycle due to accidental sensor damage. 
This damage comprises a mechanical chip in the sensor edge, where a small fragment including the bias rail around a few millimetres in width and length is physically detached; it is visible on the left edge of LS-06 as annotated in figure~\ref{fig:ppc-cracks-guides}.
Leaving the modules in dry storage for five days decreases the average bow by $22~\mu$m, indicating mechanical relaxation. 
After $[+30, -35]^\circ$C thermocycling, we find the mean bow reaches $\bar{z}_\text{bow} = 94\pm 30~\mu$m, where the average of the individual increases is $\Delta \bar{z}_\text{bow} = 27 \pm 37~\mu$m relative to $[+20, -35]^\circ$C cycling. 
The large standard deviation arises from the anomalous LS-06 value; excluding this outlier yields $\Delta \bar{z}_\text{bow} = 11 \pm 12~\mu$m.
This establishes no statistically significant change in sensor deformation during MTC from $T_\text{max} = 20^\circ$C to $30^\circ$C.

Figure~\ref{fig:bowVsTemp-nointerposer} shows substantial bow increases  when MTC reaches $T_\text{max} = 35^\circ$C and continues for $T_\text{max} = 40^\circ$C.
Table~\ref{tab:table-bow-diff-mean} shows the mean bow of $\bar{z}_\text{bow} = 68 \pm 23~\mu$m after $[+20, -35]^\circ$C cycling rises to $\bar{z}_\text{bow} = 214 \pm 13~\mu$m after $[+40, -35]^\circ$C cycling. 
This corresponds to a statistically significant bow increase of $\Delta \bar{z}_\text{bow} = 146 \pm 27~\mu$m, averaging the changes over five modules.  
It is notable that this effect is sufficiently strong to detect a bow increase for \emph{every} individual module in our data. 
This increase is interpreted as accumulated thermal stress.

We directly compare mean bow changes when raising the maximum MTC temperatures $T_\text{max}$:
\begin{alignat}{3}
    \Delta \bar{z}_\text{bow} &= 27 \pm 37~\mu\text{m}, \quad&& T_\text{max} &&= 20^\circ\text{C} \to 30^\circ\text{C},\\
    \Delta \bar{z}_\text{bow} &= 146 \pm 27~\mu\text{m}, \quad&& T_\text{max} &&= 20^\circ\text{C} \to 40^\circ\text{C}.\label{eq:non-interposer-bow-change}
\end{alignat}
This demonstrates that thermocycling above a critical temperature $T_\text{crit} \approx 35^\circ$C ``bakes in'' sensor deformations permanently, enabling \emph{ex situ} measurement outside the MTC setup.
This hysteresis is also reported in endcap modules~\cite{Salami:2025nob} and is hypothesised to arise from a glass transition of the epoxy adhesive~\cite{Carbas02012014} at $T_\text{glass} \approx 50^\circ$C.
While we lack the equipment for \emph{in situ} metrology, off-site cold metrology of an unconstrained pre-production module shows sensor corners rises by about 1~mm at $-35^\circ$C, suggesting significant thermal stress during MTC.
The mechanism of how the glass transition induces sensor deformation is not fully understood and remains an empirical observation for this paper. 

\begin{table}[]
\centering
\begin{tabular}{@{}lccccc|cc@{}}
\toprule
               & \multicolumn{7}{c}{Pre-production modules: sensor  $z_\text{bow}$ [$\mu$m]}                                              \\ 
After MTC stage & LS-01 & LS-02 & LS-03 & LS-04 & LS-06 & Mean  & Std. Dev. \\\midrule
$[+20, -35]^\circ$C & 88.2  & 95    & 60.4  & 42.9  & 51    & 67.5                      & 23.0      \\
$[+40, -35]^\circ$C & 196.4 & 223.1 & 219.3 & 203.7 & 226.6 & 213.8                     & 13.1      \\\midrule
Bow change $\Delta z_\text{bow}$ & 108.2 & 128.1 & 158.9 & 160.8 & 175.6 & 146.3                     & 27.4      \\ \bottomrule
\end{tabular}
\caption{\label{tab:table-bow-diff-mean}Selected sensor bow data for the five pre-production modules after completing MTC between $[T_\text{max}, T_\text{min}]$ temperatures from figure~\ref{fig:bowVsTemp-nointerposer}.
This table evaluates the difference between these two stages alongside the mean $\bar{z}_\text{bow}$ and standard deviation $\sigma$ over the five modules. }
\end{table}

After raising $T_\text{max}$, we lower $T_\text{min}$ beyond MTC specification, which increases module thermal stress at cold test temperatures due to CTE mismatch. 
We first attempt to lower our MTC to $T_\text{min} = -45^\circ$C.
However, the MTC inadvertently stayed at $T_\text{min} = -45^\circ$C for 12 hours due to insufficient cooling power for one chuck (reaching $-44^\circ$C) stalling the test sequence. 
Nonetheless, all chucks reach $T_\text{min} = -44^\circ$C.
Given the uncharted temperature excursion, we remove the modules and observe sensor fractures on both LS-02 and LS-06. 
After $10\times [+40, -44]^\circ$C, LS-04 sees sensor fracturing. 
After $30\times [+20, -35]^\circ$C then 30 more cycles at $[+40, -44]^\circ$C, LS-03 fractures.

We finally subject the surviving LS-01 module to 40 additional cycles between $[+45, -44]^\circ$C. 
The sensor bow of LS-01 remains above 250~$\mu$m throughout, where the highest bow value reaches 272~$\mu$m, indicating substantial stress. 
This LS-01 module nonetheless remains intact after accumulating 181 cycles.
We terminate this study here. 
Despite improved chuck flatness, we find four out of five (80\%) modules fracture after $[+40, -44]^\circ\text{C}$ cycling,
compared with one out of ten (10\%) after $[+40, -35]^\circ$C prior to this work (figure~\ref{fig:ppb2-ls06-crack}).
This precipitous rise $\sim10\% \to 80\%$ in fracturing rate reveals a critical thermal stress being exceeded. 
Fractured modules all reach $z_\text{bow}>200~\mu$m after $[+40, -35]^\circ$C MTC (table~\ref{tab:table-bow-diff-mean}), but stochastic variations in assembly and limited statistics preclude more precise explanation of specific modules failing earlier in $N_\text{cycles}$.

\FloatBarrier
\subsection{\label{subsec:postmortemfractures}Post-mortem of sensor fractures}

\begin{figure}
    \centering
    \includegraphics[width=\textwidth]{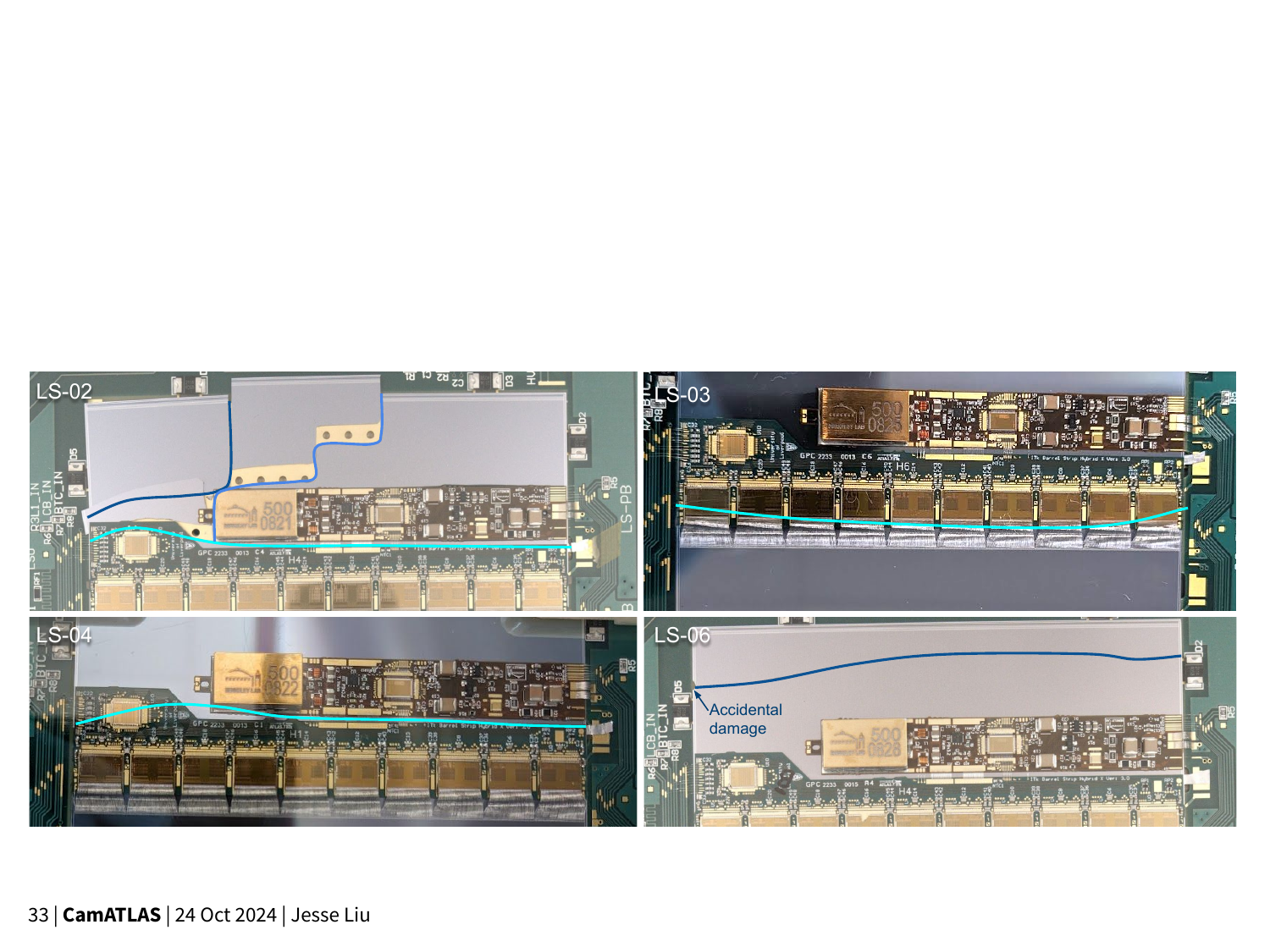}
    \label{fig:ppc-cracks-guides}
    \caption{Photos of the four pre-production modules with sensor fractures confirmed by visual inspection during this study (section~\ref{sec:non-interposer-thermocycling}). 
    Blue lines highlight the fracture locations. 
    An arrow marks the accidental mechanical damage on the LS-06 sensor during early MTC stages. 
    The metrology coordinate system starts at the top-right corner of the sensor $(x,y,z) = (0,0,0)$ with $x$ leftwards, and $y$ downwards, and $z$ out of page.
    }
    \label{fig:cracks-guides}
\end{figure}

We now examine the modules post mortem to discuss similarities and differences in fracture morphology.
We identify their locations by visual inspection in figure~\ref{fig:cracks-guides} for the fractured modules:
\begin{itemize}
    
    \item \textbf{Similarities in fracture morphology}. 
    A common morphology observed on all modules is a single \emph{primary fracture} that completely bifurcates the sensor edge to edge.
    The fracture direction is parallel to the longer edges of the PCBs.
    This supports the hypothesis of a common mechanism for fracturing. 
    It suggests thermal stress concentrated parallel to the long PCB edge. 
    The primary fracture for LS-02 and LS-04 are especially similar, situated in the narrow gap between powerboard and hybrid, becoming concealed as it turns under the hybrid flex by the HCCStar chip on the left.
    While it is difficult to directly observe beneath the flex without peeling off the PCB, the primary fracture likely follows the glue boundary. 

    The LS-03 primary fracture follows the long edge of the hybrid but is largely concealed by the flex on the left/right-most edges and the front-end bonds of the central hybrid chips.
    The location being under front-end bonds directly reproduces the location of the fracture prior to this study (figure~\ref{fig:ppb2-ls06-crack}).
    A small secondary sensor fragment is observed on the leftmost edge. 
    
    \item \textbf{Differences in fracture morphology}.
    The LS-02 sensor distinctly fractures into multiple pieces, where we visually identify five major segments. 
    We refer to the fractures that do not propagate from one edge to the opposite edge as \emph{secondary fractures}.
    We observe that these develop perpendicular to the primary fracture left of the DC-DC converter shield box on the powerboard.
    One secondary fracture (medium blue) follows the powerboard glue edge with three near-right-angle kinks turning to the sensor top edge.
    The secondary fracture (dark blue) arises just above the primary fracture by the hybrid before turning perpendicularly at the top-left corner of the powerboard shield box to the sensor top edge, which does not trace a glue boundary. 

    The LS-06 sensor suffers chip damage at the left edge near the top edge due to accidental handling after the first $[+25, -35]^\circ$C MTC.
    Unlike the other primary fractures, the location is not adjacent to the PCB edge, but appears seeded from that chip location across to the opposite side.
    This suggests sensor damage elevates localised stress and fracturing vulnerability.

\end{itemize}

\begin{table}[]
\centering
\begin{tabular}{@{}ccc@{}}
\toprule
Material & Modulus [GPa] & CTE $\left[10^{-6}~\text{K}^{-1}\right]$ \\ \midrule
Silicon  & 160           & 2.6          \\
Copper   & 120           & 16.7         \\
Kapton   & 2.5           & 20           \\
Epoxy    & 3.1           & 60           \\ 
SE-4445  & 0.001         & ---          \\\bottomrule
\end{tabular}
\caption{\label{tab:module-modulus-cte}Indicative modulus and coefficients of thermal expansion (CTE) for module materials. 
Circuit board flexes comprise a composite stack of copper and Kapton. }
\end{table}

These fracture locations are qualitatively compatible with finite-element analysis (FEA) simulation modelling thermal stress~\cite{itk-strips-cracks-simulation}.
Flexes comprise a stackup with alternating layers of copper and Kapton, where cooling causes the copper to shrink more than the sensor due to mismatched CTE (table~\ref{tab:module-modulus-cte}).
Suction cups in the handling frame induce a long lever arm that exerts stress at the sensor interface with the epoxy (figure~\ref{fig:module-ppb-stackup}).
The epoxy glass transition induces a sensor bow hysteresis, causing stress to accumulate~\cite{Salami:2025nob}. 
Levelling thermal chucks should reduce apparatus-induced stress  (subsection~\ref{subsec:Tchuck}), but this mitigation is evidently insufficient to prevent sensor deformation and fracturing.
This indicates that thermal stress is intrinsic to pre-production modules~\cite{ATLAS-TDR-25}, which motivates design modifications for production.

\FloatBarrier
\section{\label{sec:interposers}Sensor deformation with stress-mitigating interposer}

\begin{figure}
    \centering
    \begin{subfigure}{0.25\textwidth}
    \centering
    \includegraphics[width=\textwidth]{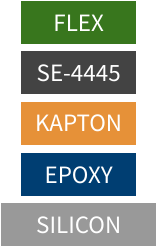}
    \caption{\label{fig:interposer-stackup-schematic}Schematic stackup}
    \end{subfigure}
    \begin{subfigure}{0.73\textwidth}
    \centering
    \includegraphics[width=\textwidth]{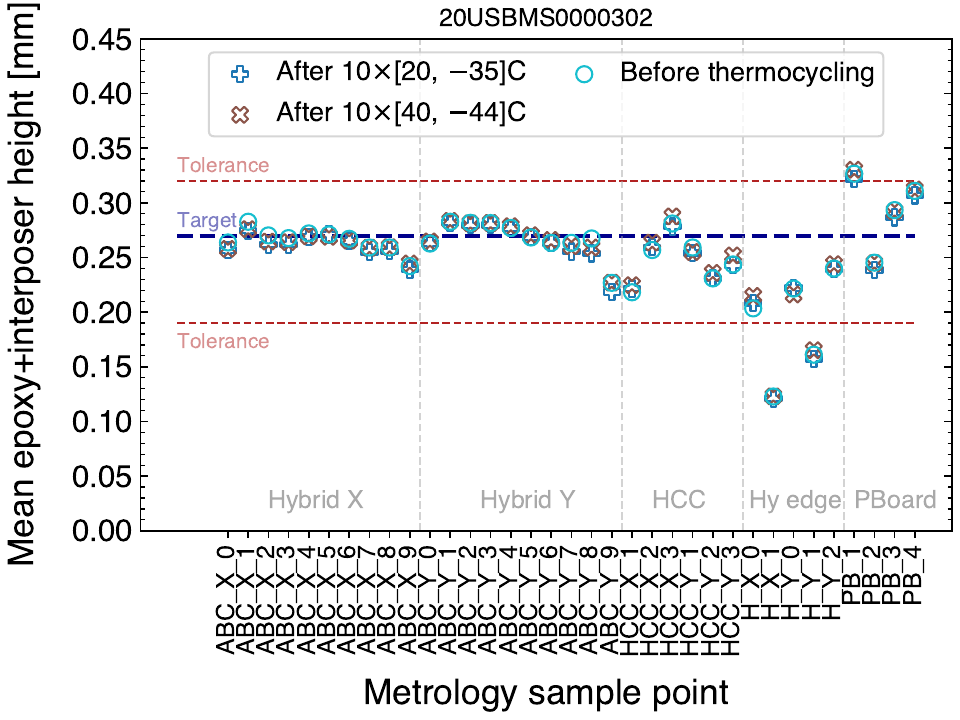}
    \caption{\label{fig:interposer-glue-height-metrology}Combined epoxy plus interposer height metrology}
    \end{subfigure}
    \caption{(a) Schematic of the Kapton-silicone interposer stackup between the flex and epoxy introduced in Refs.~\cite{itk-strips-cracks-simulation,Fortman:2926236}. 
    (b) Heights between flex and sensor for the SS-04 module, 
    whose target (thick dashed line) is the sum of $120~\mu$m epoxy, 50~$\mu$m Kapton, 100~$\mu$m SE-4445.
    Heights are sampled across hybrid microchips (ABC, HCC), hybrid edges (Hy edge), and powerboard (PBoard) positions defined by common routines~\cite{Tishelman-Charny:2024yys}.
    Data are taken before MTC ($\circ$ marker), and those after $10\times [20, -35]^\circ$C ($+$ marker) and $10\times [40, -44]^\circ$C ($\times$ marker).
    Error bars are suppressed for visual clarity but typical ranges are 0.01--0.03~mm.}
\end{figure}

Simulation studies~\cite{itk-strips-cracks-simulation} suggest an order-of-magnitude reduction in stress by adding Kapton-silicone interposers. 
Due to its low modulus, the silicone gel reduces mechanical stress by decoupling the thermal deformation of the flexes from that of the sensor during cooling~\cite{Matayabas:2005,larson-dow}.
Figure~\ref{fig:interposer-stackup-schematic} schematically shows the flex-epoxy-sensor stackup augmented with 50 $\mu$m of Kapton film and 100 $\mu$m of SE-4445 silicone gel, whose concept is introduced in Refs.~\cite{itk-strips-cracks-simulation,Fortman:2926236}. 
The Kapton film is required to ensure sufficient adhesion given it is found that the epoxy does not adhere well to silicone. 
These two materials are already validated for use in the ITk detector because SE-4445 attaches modules onto local supports, whose electrical services supplied by bus tapes are made of Kapton~\cite{Weidberg:2025dmx}. 
Introducing these extra thin layers across all modules has a negligible impact on ITk material budget. 
This motivates us to prototype interposers in our laboratory by developing in-house designs, ad hoc assembly procedures, tests of engineering viability, and probes for reduced sensor deformation.

\subsection{\label{sec:interposers-assembly}Ad hoc attachment of interposer to module}

This subsection introduces the early ad hoc procedure\footnote{This is referred to as the ``cut and roll method'' using tooling repurposed from the fine arts community~\cite{Essdee-ink-roller}.} to attach silicone gel and Kapton film onto flexes.   
Test pieces are assembled to verify the viability of the method to achieve the desired glue thickness and uniformity.

\begin{figure}
    \centering
    \begin{subfigure}{0.30\textwidth}
    \centering
    \includegraphics[height=4.1cm]{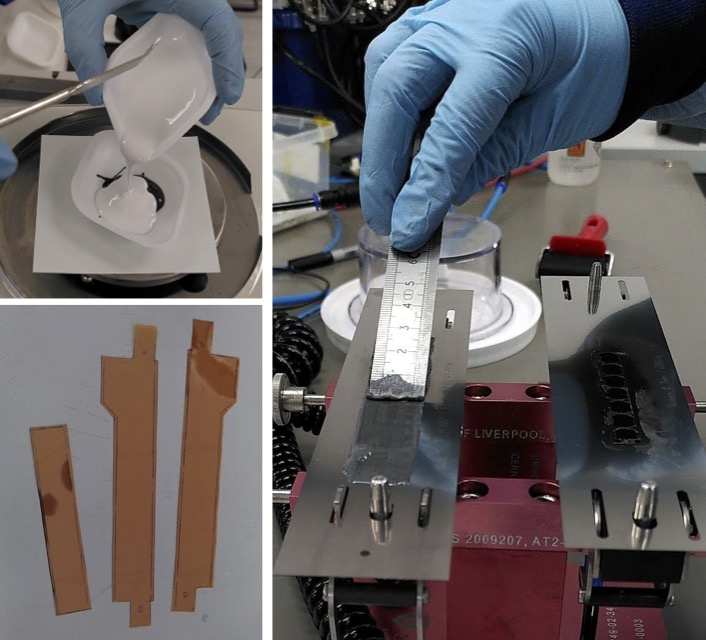}
    \caption{Prepare interposers}
    \end{subfigure}%
    \begin{subfigure}{0.34\textwidth}
    \centering
    \includegraphics[height=4.1cm]{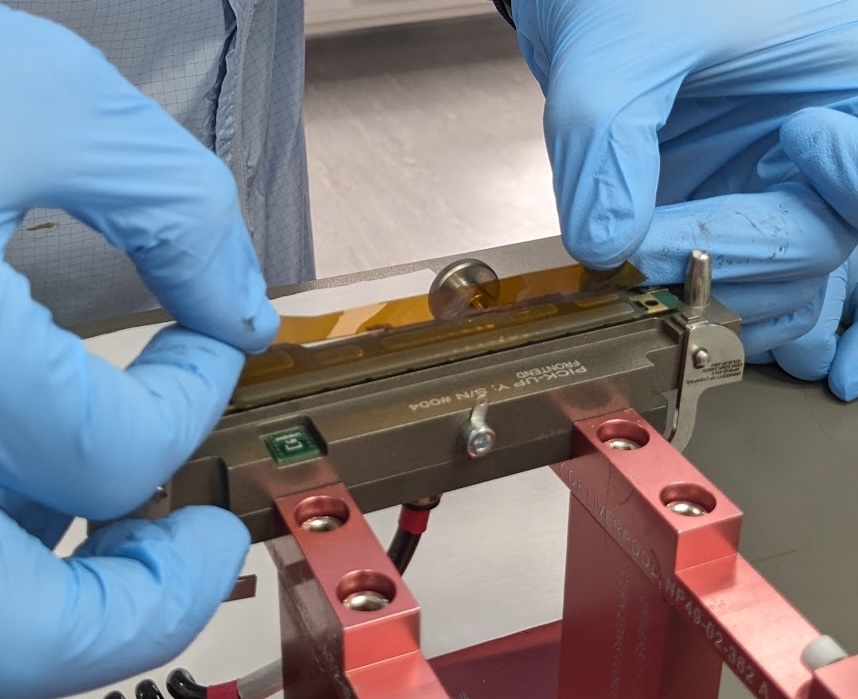}
    \caption{Deposit Kapton on SE-4445}
    \end{subfigure}%
    \begin{subfigure}{0.34\textwidth}
    \centering
    \includegraphics[height=4.1cm]{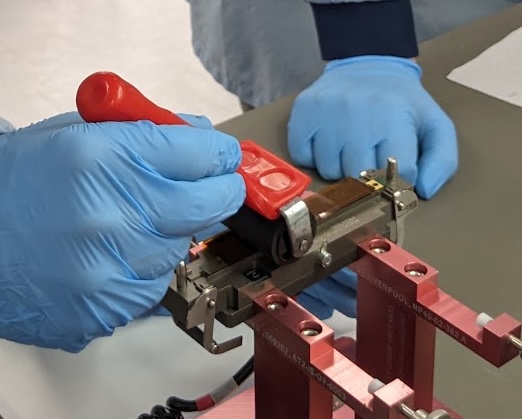}
    \caption{Roll for full coverage}
    \end{subfigure}
    \caption{\label{fig:interposer-photos}Ad hoc procedure for applying the first prototype Kapton-silicone interposers onto flexes. 
    (a) This shows the preparation of the two-part SE-4445 gel (top left), manually cut Kapton interposers (bottom left), stencilling SE-4445 onto the rear of hybrids.
    (b) This shows the deposition of 50~$\mu$m Kapton film (orange) onto the 100~$\mu$m layer of SE-4445 silicone gel (gray) stencilled onto the hybrid flex. 
    (c) This shows the use of a print roller~\cite{Essdee-ink-roller} to apply pressure onto the Kapton-silicone interposer to ensure full SE-4445 coverage before leaving overnight to cure.}
\end{figure}

Figure~\ref{fig:interposer-photos} displays the early ad hoc procedure for attaching interposers onto flexes before assembling into a module. 
The SE-4445 is a two-part mixture that is hand mixed in a 1:1 ratio by weight. 
It is found that SE-4445 contamination on sensors induces undesirable early HV breakdown. 
The Kapton film is therefore oversized by 1~mm all-around to ensure full flex coverage while avoiding SE-4445 spilling onto the sensor surface.
We initially use a mechanical hybrid and powerboard to test this procedure before using electrical counterparts, verifying the target glue thickness is achievable while ensuring satisfactory glue coverage.
The Kapton film is manually cut to shape using scissors.
Existing stencils (designed to deposit the F112 epoxy glue layer) are repurposed to deposit the SE-4445 onto the flexes using a steel ruler as a squeegee. 
After lifting the stencil, the operator deposits the Kapton film manually onto the SE-4445. 
An Essdee ink roller~\cite{Essdee-ink-roller} is procured to apply pressure and ensure full coverage of SE-4445 by visual inspection. 
Upon SE-4445 compression, the fill factor of the repurposed stencil approximately achieves the target thickness that is later verified by metrology. 
We then use existing tooling shimmed to account for the added $150~\mu$m thickness for assembly onto a sensor.
A pre-production ATLAS18SS sensor is used with good high-voltage characteristics.

\subsection{\label{sec:interposers-first-prototype-qc}Quality control of module with ad hoc interposer}

The resulting interposer module is called SS-04 (figure~\ref{fig:assembled-SS-module}). 
Any modified module design must undergo comprehensive validation. 
It must not introduce adverse side effects such as delamination from insufficient adhesion, excess operating temperatures due to insufficient thermal conduction, and excess leakage current and electrical noise. 
It should not exacerbate a long-standing \emph{cold noise} problem~\cite{Dyckes:2891402}, where electronics vibrations cause excess readout noise for SS modules at $-35^\circ$C. 
We perform and present detailed quality control test data for the SS-04 interposer module:
\begin{itemize}
    \item \textbf{Glue height metrology}. 
    Figure~\ref{fig:interposer-glue-height-metrology} shows the glue height metrology for SS-04. 
    Circle markers show the target heights expected for the interposer stackup.
    This verifies our ad hoc stencilling procedure can control the deposition of the SE-4445, Kapton, and epoxy. 
    A small minority of points fall outside tolerance, but has negligible impact on downstream assembly such as wire bonding so is deemed unproblematic. 
    Cross markers show the glue height measurements after MTC at two temperature ranges. 
    The differences are negligible, confirming no undesirable delamination.

    \item \textbf{On-module temperatures}. 
    Figure~\ref{fig:interposer-amac-temperatures} shows the on-module temperatures for the powerboard and hybrid NTCs during $10\times[+20, -35]^\circ$C MTC.
    These temperatures are read out by the Autonomous Monitoring and Control (AMAC) microchip~\cite{Gosart:2023pcl}.
    There are slight increases of around $4^\circ$C relative to non-interposer modules.
    This is compatible with expectations from FEA thermal simulation based on Ref.~\cite{Beck:2020qme}.
    Consistency of AMAC temperatures is also an indirect test for delamination during MTC, where flexes or interposers detaching from the epoxy would reduce cooling to electronics thus elevating temperatures. 
    No readings exceed their operational temperatures due to insufficient thermal conduction from the additional interposer materials.

\begin{figure}
    \centering
    \includegraphics[width=\linewidth]{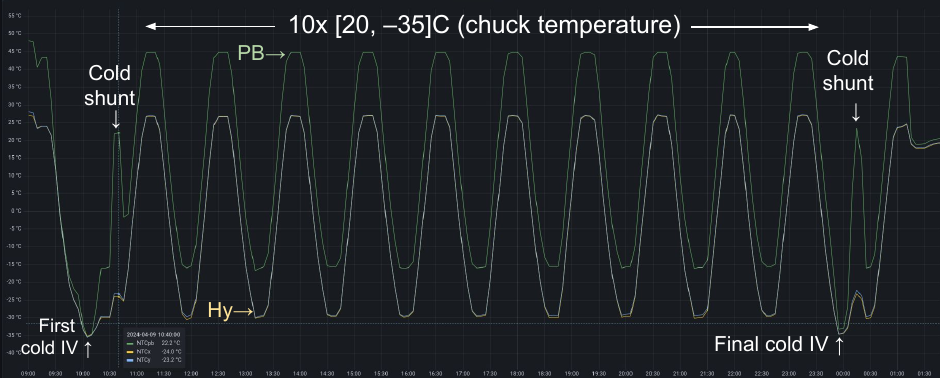}
    \caption{\label{fig:interposer-amac-temperatures}On-module temperatures during $10\times [20, -35]^\circ$C thermocycling of the SS-04 interposer module.
    Displayed are the AMAC readings for the hybrid X (yellow), hybrid Y (blue), and powerboard (green). 
    The indicated first and final current-voltage (IV) scans at cold $-35^\circ$C chuck temperatures with readout chips switched off, reducing heat load.
    The ``cold shunt'' tests see elevated power beyond nominal operating conditions, increasing heat load. }
\end{figure}

    \item \textbf{Leakage current}. 
    Sudden exponential growth in leakage currents is the signature for HV breakdown.
    This is observed in $-35^\circ$C testing prior to this study,
    which is hypothesised to arise from adverse sensor deformation. 
    It is also vital to verify if interposer materials placed near sensors induce HV breakdown~\cite{Helling:2019zea}. 
    Figure~\ref{fig:interposer-prototype-iv} shows the leakage current vs voltage scans for SS-04.
    Before module bonding, direct sensor scans reach 700~V bias voltage.
    Following bonding, we use the AMAC to measure leakage currents up to 550~V in the MTC routine before and after each of three MTC temperature ranges: $10\times[+20, -35]^\circ$C, $10\times[+20, -44]^\circ$C, and $10\times[+40, -44]^\circ$C. 
    Moreover, 
    the AMAC leakage currents is stable during continuous monitoring throughout $10\times[+20, -44]^\circ$C, and $10\times[+40, -44]^\circ$C MTC.
    No early breakdown signatures are observed.

\begin{figure}
    \centering
    \begin{subfigure}{0.43\textwidth}
    \centering
    \includegraphics[width=\textwidth]{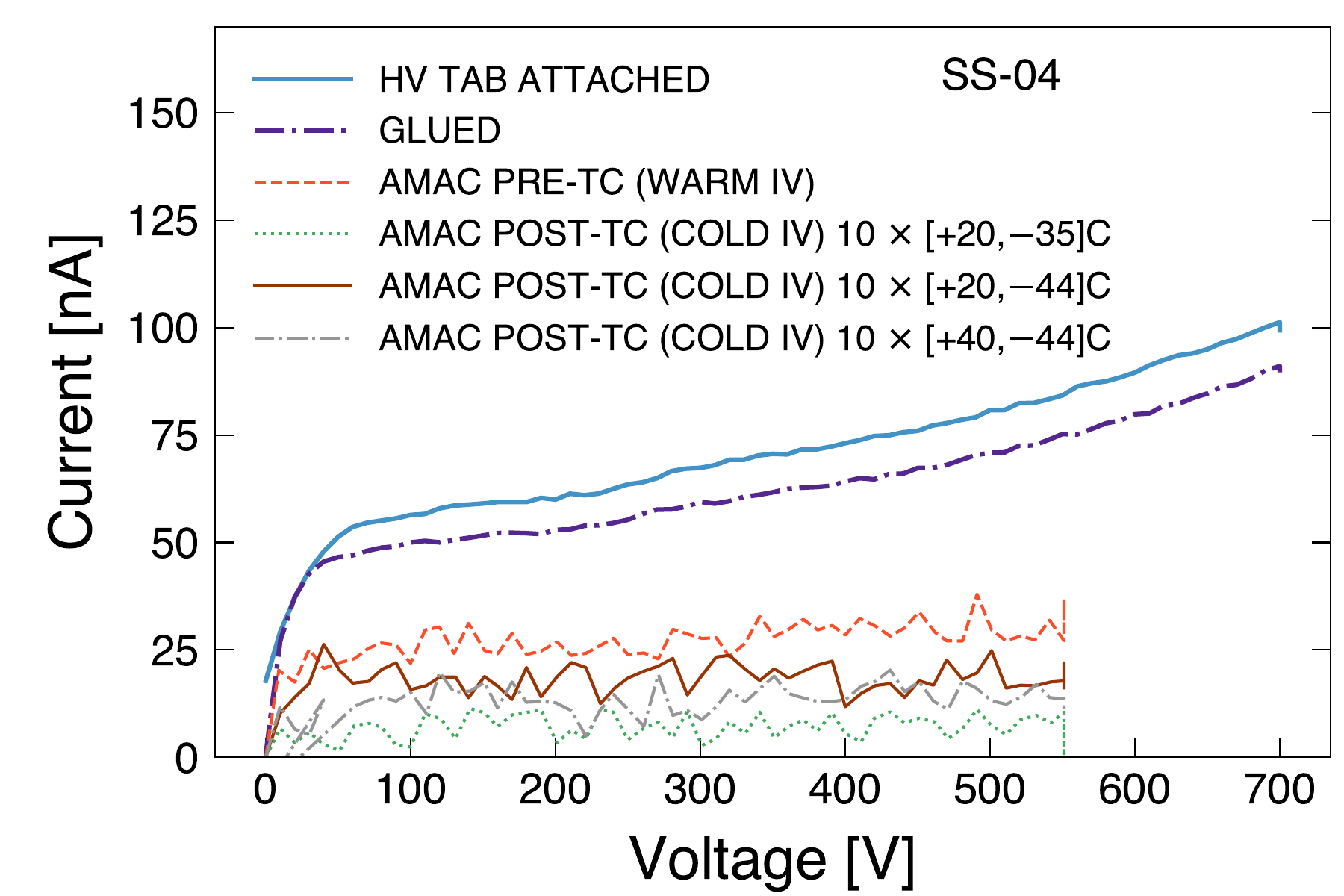}
    \caption{\label{fig:interposer-prototype-iv}Current-voltage (IV) scans}
    \end{subfigure}
    \begin{subfigure}{0.56\textwidth}
    \includegraphics[width=\textwidth]{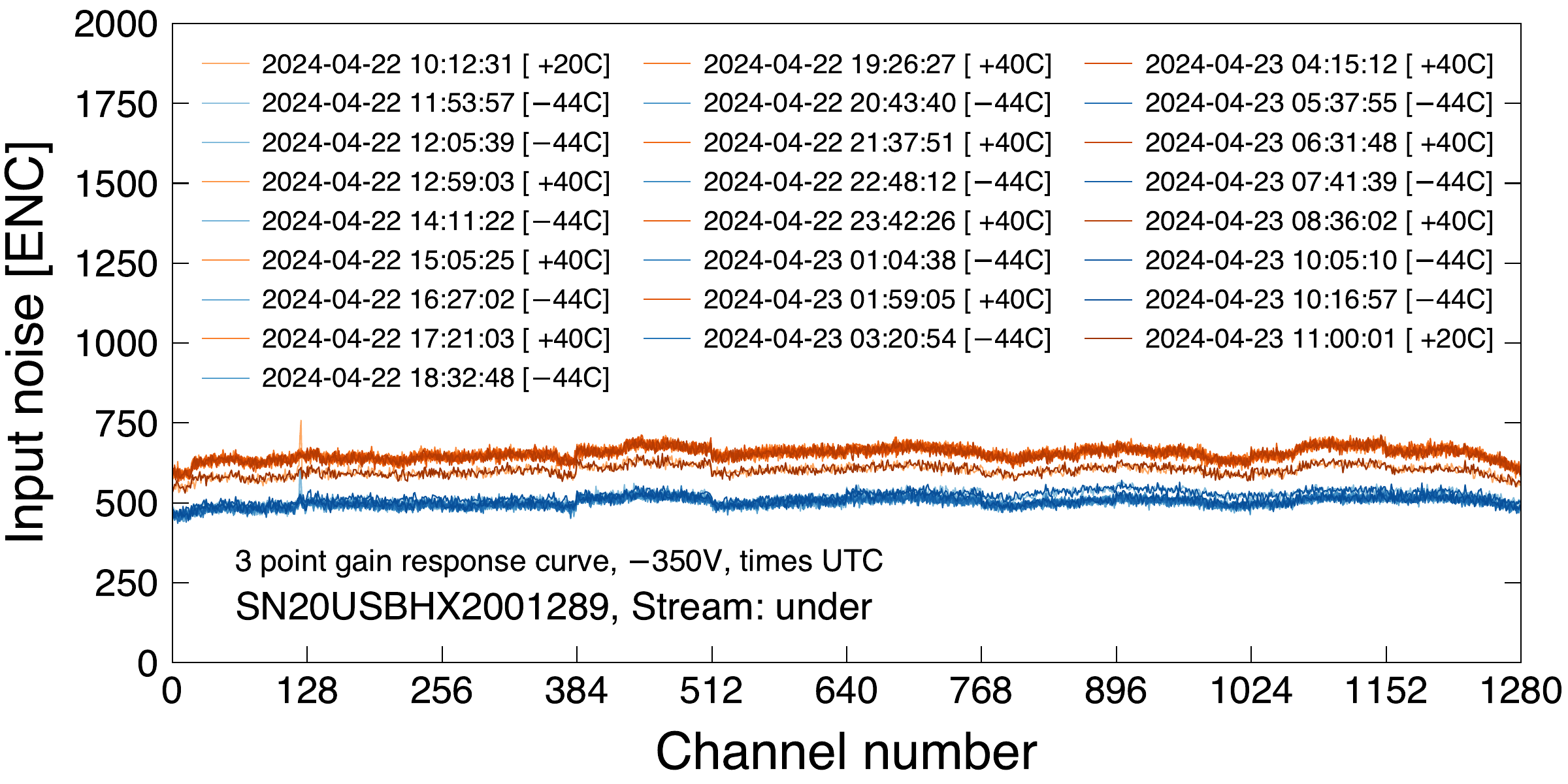}
    \caption{\label{fig:interposer-prototype-input-noise}Per-channel input noise}
    \end{subfigure}
    \caption{
    Electrical test results for first interposer SS-04 module. 
    (a) 
    Current--voltage scans with (without) `AMAC' occur before bonding up to 550~V (700~V).
    Pre-TC (warm IV) occurs just before thermocycling starts at $+20^\circ$C.
    Post-TC (cold IV) scans occur at the $T_\text{min}$ temperature after the noted $N_\text{cycles} \times [T_\text{max}, T_\text{min}]$ run.
    (b) Every three-point gain input noise result during $10\times[+40, -44]^\circ$C thermocycling is shown for the hybrid X under stream.
    Oranges (blues) indicate tests above (below) $0^\circ$C; lighter (darker) lines show earlier (later) testing.} 
\end{figure}

    \item \textbf{Readout input noise}. 
    During MTC, module electrical tests are performed at $T_\text{min}$ and $T_\text{max}$ using the community-developed ITk Strip Data Acquisition (ITSDAQ) software.
    These include threshold ``VT50'' scans to determine the \emph{output noise} where the number of hits reaches 50\% of the triggers, after which ``three point gain'' tests inject [0.5, 1.0, 1.5]~fC of charge to measure the \emph{gain}~\cite{ATLAS:2020ize}.
    
    The ratio of output noise and gain determines the \emph{input noise} for each channel.  
    Two tests at $+20^\circ$C occur before and after the main MTC sequence as baseline cross-checks. 
    Pre-production SS modules typically exhibit anomalous excesses of input noise during $-35^\circ$C testing, which is the hallmark of cold noise~\cite{Dyckes:2891402}.
    This is not observed for SS-04.
    The severity of cold noise increases with (i) lower temperatures and (ii) increasing the power load on the DC-DC converter on the powerboard as described in Ref.~\cite{Dyckes:2891402}. 
    To test (i), we perform $10\times[+40, -44]^\circ$C testing and results for one set of readout channels. 
    To test (ii), a ``cold shunt test'' occurs immediately after the first and last cold test in the standard MTC sequence, which temporarily shunts the current on the front-end chips to increase its power draw and thus load on the DC-DC converter. 

    No anomalous noise structures are observed at $-44^\circ$C for any test across all 5120 channels. 
    Figure~\ref{fig:interposer-prototype-input-noise} shows the results of all these tests for half the channels of one hybrid.
    All channels exhibit lower input noise by around 100~Equivalent Noise Charge (ENC) at $-44^\circ$C than at $+20^\circ$C and $+40^\circ$C~\cite{Cormier:2021oog}. 
    Negligible change in input noise between cycles verifies the stability of results. 
    This also contrasts with anomalous noise structures in pre-production SS modules during $-35^\circ$C testing~\cite{Dyckes:2891402}. 
    Ad hoc interposer attachment thus not only lacks adverse impact on electrical noise, but even ameliorates the pre-production problem of cold noise.
    The precise mechanism of cold noise mitigation is not fully understood, but is hypothesised to arise from the low modulus of SE-4445 dissipating vibrations. 

\begin{figure}
    \centering
    \begin{subfigure}{0.24\textwidth}
    \centering
    \includegraphics[trim={2.9cm 0.1cm 0.8cm 0.1cm},clip,width=\textwidth]{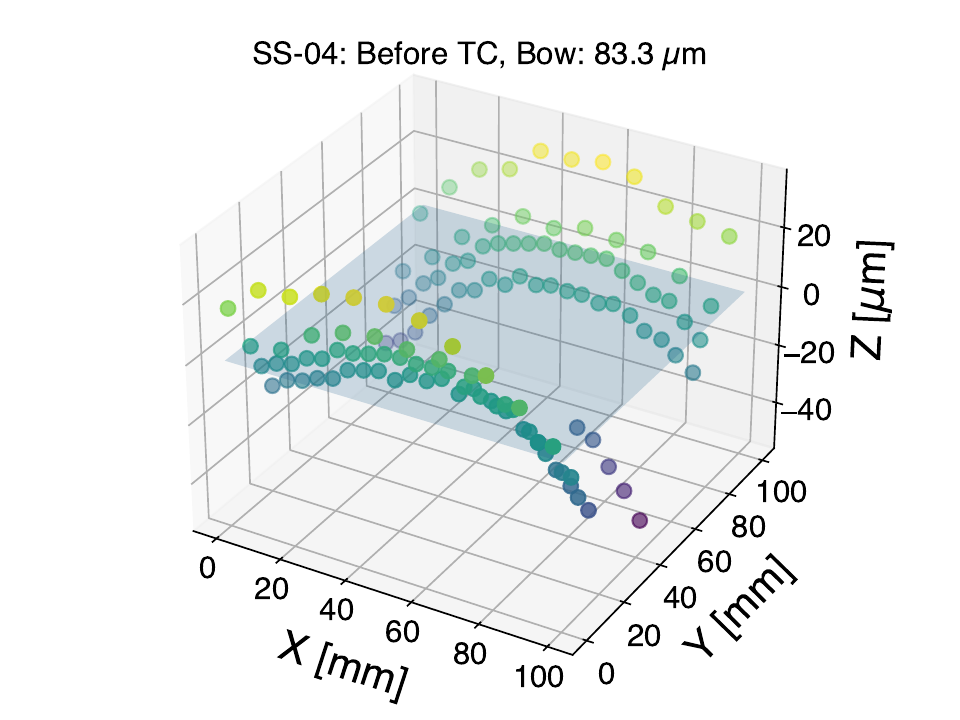}
    \caption{Before MTC}
    \end{subfigure}%
    \begin{subfigure}{0.24\textwidth}
    \centering
    \includegraphics[trim={2.9cm 0.1cm 0.8cm 0.1cm},clip,width=\textwidth]{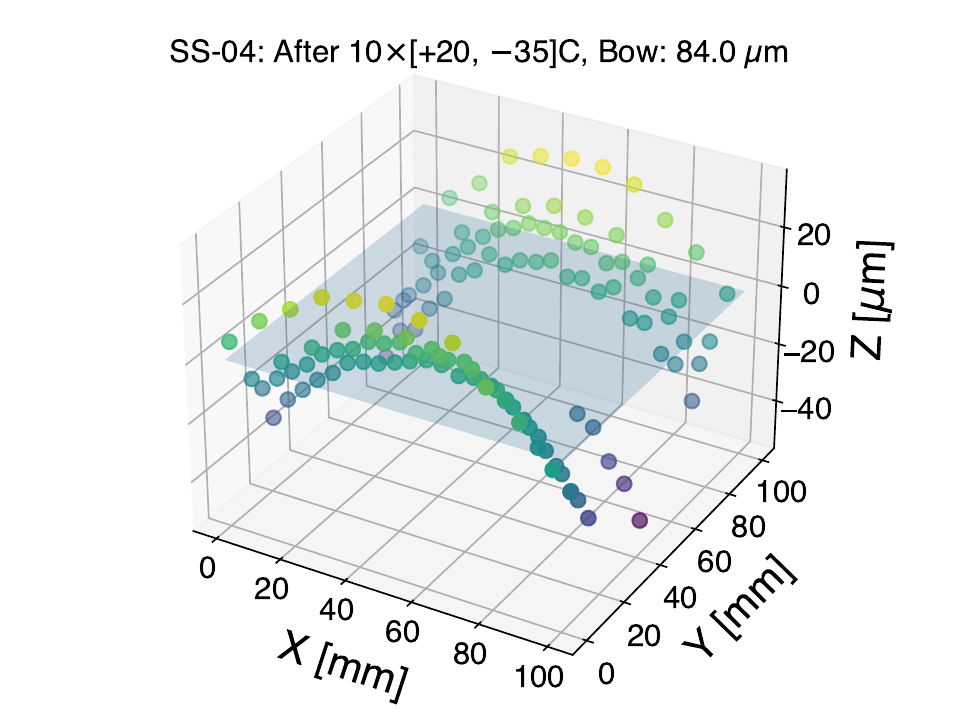}
    \caption{ $10\times[+20, -35]^\circ$C}
    \end{subfigure}%
    \begin{subfigure}{0.24\textwidth}
    \centering
    \includegraphics[trim={2.9cm 0.1cm 0.8cm 0.1cm},clip,width=\textwidth]{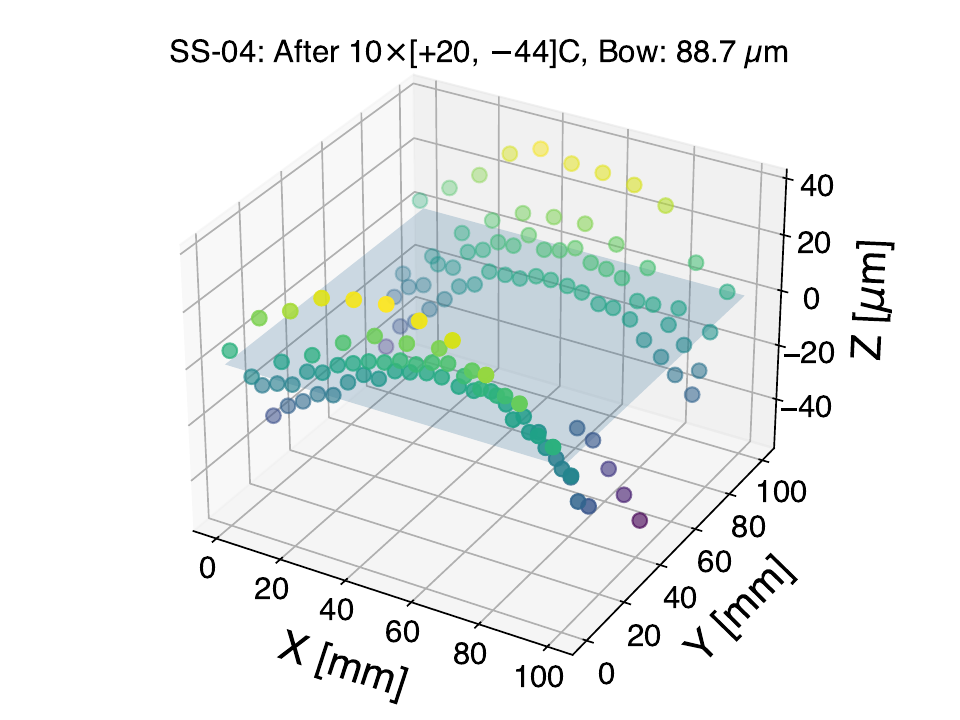}
    \caption{ $10\times[+20, -44]^\circ$C}
    \end{subfigure}
    \begin{subfigure}{0.24\textwidth}
    \centering
    \includegraphics[trim={2.9cm 0.1cm 0.8cm 0.1cm},clip,width=\textwidth]{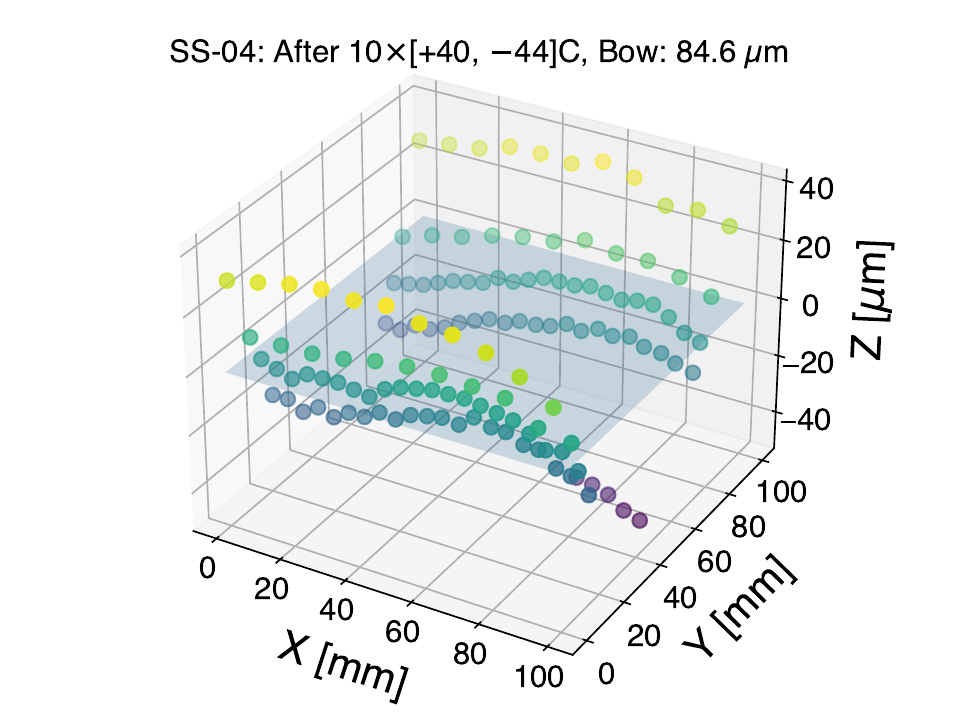}
    \caption{ $10\times[+40, -44]^\circ$C}
    \end{subfigure}
    \caption{\label{fig:ss04-bow-metrology}Sensor height metrology of SS-04 before and after thermocycling through the $N_\text{cycles} \times[T_\text{max}, T_\text{min}]$ temperatures indicated.}
\end{figure}

    \item \textbf{Sensor bow metrology}.
    The sensor bow $z_\text{bow}$ of SS-04 is measured before and after module thermocycling for the following temperature ranges:
    \begin{align}
    \text{Before } 10\times[+20, -35]^\circ\text{C}: \quad& z_\text{bow} = 83.7~\mu\text{m},\nonumber\\
        \text{After } 10\times[+20, -35]^\circ\text{C}: \quad& z_\text{bow} = 84.0~\mu\text{m},\nonumber\\
        \text{After } 10\times[+20, -44]^\circ\text{C}: \quad& z_\text{bow} = 88.7~\mu\text{m},\nonumber\\
        \text{After } 10\times[+40, -44]^\circ\text{C}: \quad& z_\text{bow} = 84.6~\mu\text{m}.
     \label{eq:cam-ppb2-ss-04-bow}
    \end{align}
    These sensor bow values change by no more than $5~\mu$m after $10\times[+40, -44]^\circ\text{C}$ cycling. 
    This alone is a significant result compared with \emph{every} non-interposer module seeing its bow increase by a mean of $146\pm 27~\mu$m after $T_\text{max} = +40^\circ$C MTC (table~\ref{tab:table-bow-diff-mean}).
    Figure~\ref{fig:ss04-bow-metrology} shows the sensor shape metrology for these four stages has negligible qualitative changes. 
    This reduced sensor deformation indicates mitigated thermal stress.

\end{itemize}

Together, these detailed tests of this SS-04 interposer module with no adverse side effects are promising.
The additional observations of mitigated cold noise and no significant sensor bow change is remarkable.
This motivates further interposer modules to test reproducibility and reliability. 
Following a total of 30 thermocycles after the $10\times [+40, -44]^\circ$C run, SS-04 is permanently sent away for off-site test beam and irradiation studies~\cite{Ravotti:2014uda,Gkotse:2023dff}.

To continue in-house QA studies, we additionally construct one long-strip (LS-12) and another short-strip (SS-05) interposer module. 
For these two modules, we attach the silicone gel and Kapton film to flexes using module assembly tooling with shims to improve procedural reliability.
The vacuum tooling is repurposed with shims to control consistency of the SE-4445 target height during curing. 
Once assembled, these modules undergo extended QA thermocycling to probe their mechanical resilience. 
We proceed directly to $T_\text{min}=-44^\circ$C after the first ten thermocycles to expedite thermal stress.
Due to epoxy squeeze-out over the sensor bias rail, SS-05 suffers early breakdown at $\approx 200$~V so undergoes MTC with powered LV electronics but disconnected HV.

\subsection{\label{subsec:evidence-mitigated-stress}Reduced sensor deformation and thermal stress}
\FloatBarrier

\begin{figure}
    \centering
    \includegraphics[width=\textwidth]{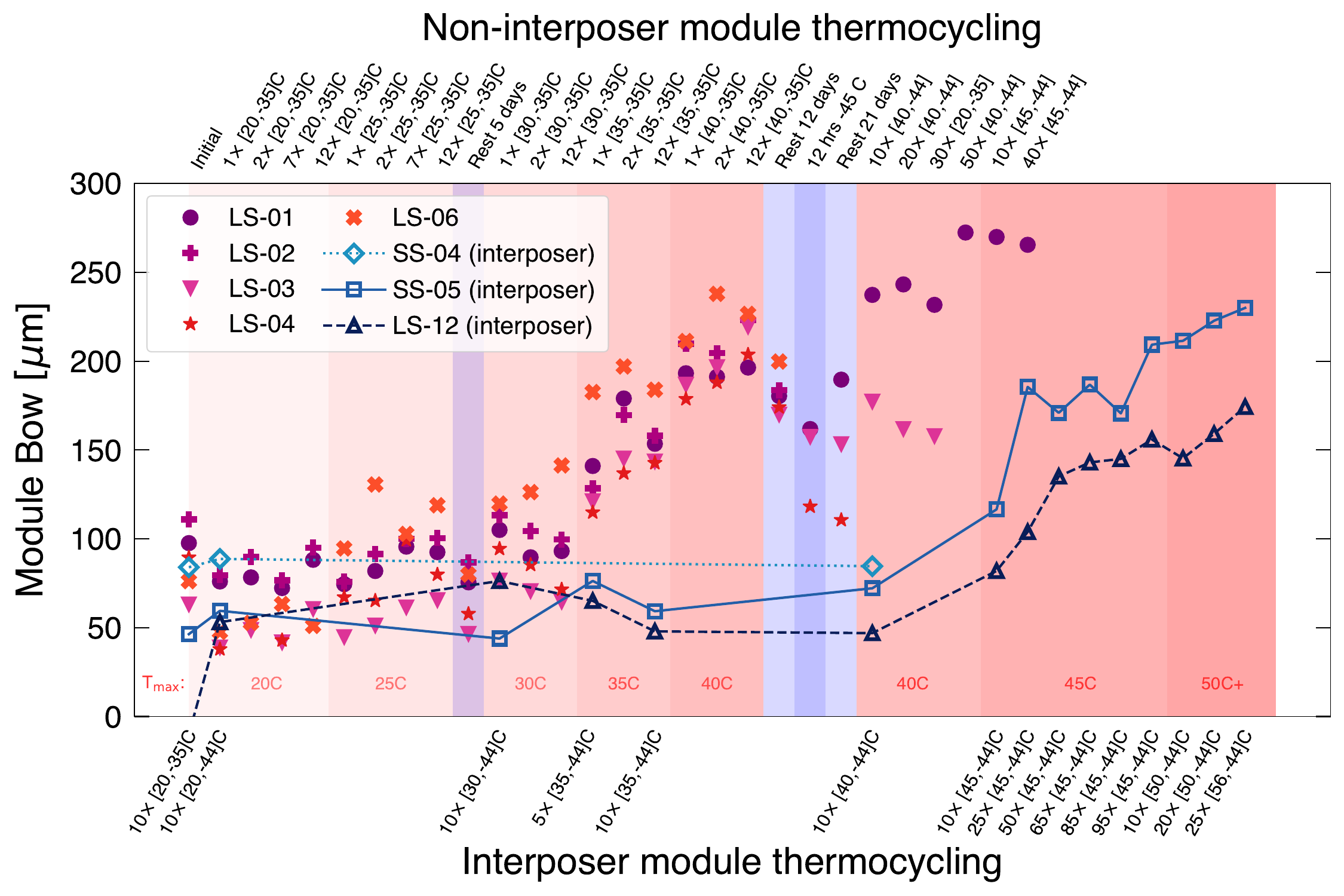}
    \caption{\label{fig:bowVsTemp} 
    Summary of sensor bow with thermocycling step for five pre-production modules (filled markers, upper axis) and three interposers modules (hollow markers with lines, lower axis).
    Axis labels show $N\times$ accumulated cycles between $[T_\text{max}, T_\text{min}]$ temperatures.
    Each marker corresponds to a subsequent bow measurement. 
    Red shading shows the maximum temperature $T_\text{max}$ labelled on the plot. 
    Blue shading shows bow metrology after resting pre-production modules in dry storage for the number of days in the upper axis. 
    }
\end{figure}

We now analyse in detail how sensor deformation of interposer modules differs from non-interposer counterparts.
Figure~\ref{fig:bowVsTemp} shows a systematic divergence at $T_\text{max} = +35^\circ$C and $+40^\circ$C MTC: bow trends increase in all five non-interposer modules (filled markers) while no interposer ones (hollow markers) see significant changes.  
While there are stochastic variations, \emph{every} module in each interposer category exhibits the same bow trends, indicating highly correlated effects.
This enables our modest sample size of modules to detect such systematic changes when introducing interposers. 

\begin{table}[]
\centering
\begin{tabular}{@{}lccc|cc@{}}
\toprule
               & \multicolumn{5}{c}{Interposer modules: sensor bow  $z_\text{bow}$ [$\mu$m]}              \\
After MTC stage & SS-04 & LS-12 & SS-05 & Mean & Std.\ Dev. \\
\midrule
$[+20, -44]^\circ$C & 88.7  & 53.1 & 59.5 & 67.1                      & 19.0      \\
$[+40, -44]^\circ$C & 84.6  & 46.9 & 72.2 & 67.9                      & 19.2      \\
\midrule
Bow change $\Delta z_\text{bow}$   & $-4.1$& $-6.2$& 12.7 & 0.8                       & 10.4     \\
\bottomrule
\end{tabular}
\caption{\label{tab:table-bow-diff-mean-interposer}Selected sensor bow data for the three interposer modules after completing MTC between $[T_\text{max}, T_\text{min}]$ temperatures from figure~\ref{fig:bowVsTemp}.
This table evaluates the difference between these two stages alongside the mean $\bar{z}_\text{bow}$ and standard deviation $\sigma$ over the three modules.}
\end{table}

Table \ref{tab:table-bow-diff-mean-interposer} quantifies this more precisely with selected interposer module data. 
The sensor bow averaged over these three modules is $67 \pm 19~\mu$m after $10\times [+20, -44]^\circ$C and remains unchanged  $68 \pm 19~\mu$m after $10\times [+40, -44]^\circ$C.
Evaluating individual bow changes $\Delta z_\text{bow}$ then averaging yields $\Delta \bar{z}_\text{bow} = 1\pm 10~\mu\text{m}$. 
We directly contrast this with the non-interposer module results:
\begin{alignat}{4}
    \text{Non-interposer}: \quad& \Delta \bar{z}_\text{bow}^\text{no-int} &&= 146\pm 27~\mu\text{m}, \quad&& T_\text{max} = 20^\circ\text{C} \to 40^\circ\text{C}, \quad&& (\text{table}~\ref{tab:table-bow-diff-mean}), \\
    \text{Interposer}: \quad & \Delta \bar{z}_\text{bow}^\text{int} &&= 1\pm 10~\mu\text{m}, \quad&& T_\text{max} = 20^\circ\text{C} \to 40^\circ\text{C}, \quad&& (\text{table}~\ref{tab:table-bow-diff-mean-interposer}).
\end{alignat}
A simple measure of their statistical incompatibility is $\left(\Delta \bar{z}_\text{bow}^\text{no-int} - \Delta \bar{z}_\text{bow}^\text{int}\right)/\sqrt{\sigma_\text{no-int}^2 + \sigma_\text{int}^2} = 5.0$, normalised to the quadrature sum of standard deviations $\sigma$. 
This quantifies a significant reduction in sensor deformation during MTC from $T_\text{max} = 20^\circ$C to $40^\circ$C. 
By contrast, every non-interposer module exhibits consistent bow increases and all but one fractures (section~\ref{subsec:bowvscycling}).
Together, these results are interpreted as evidence that interposers are a promising stress mitigation strategy.

\begin{figure}
    \centering
    \begin{subfigure}{0.49\textwidth}
    \centering
    \includegraphics[width=\textwidth]{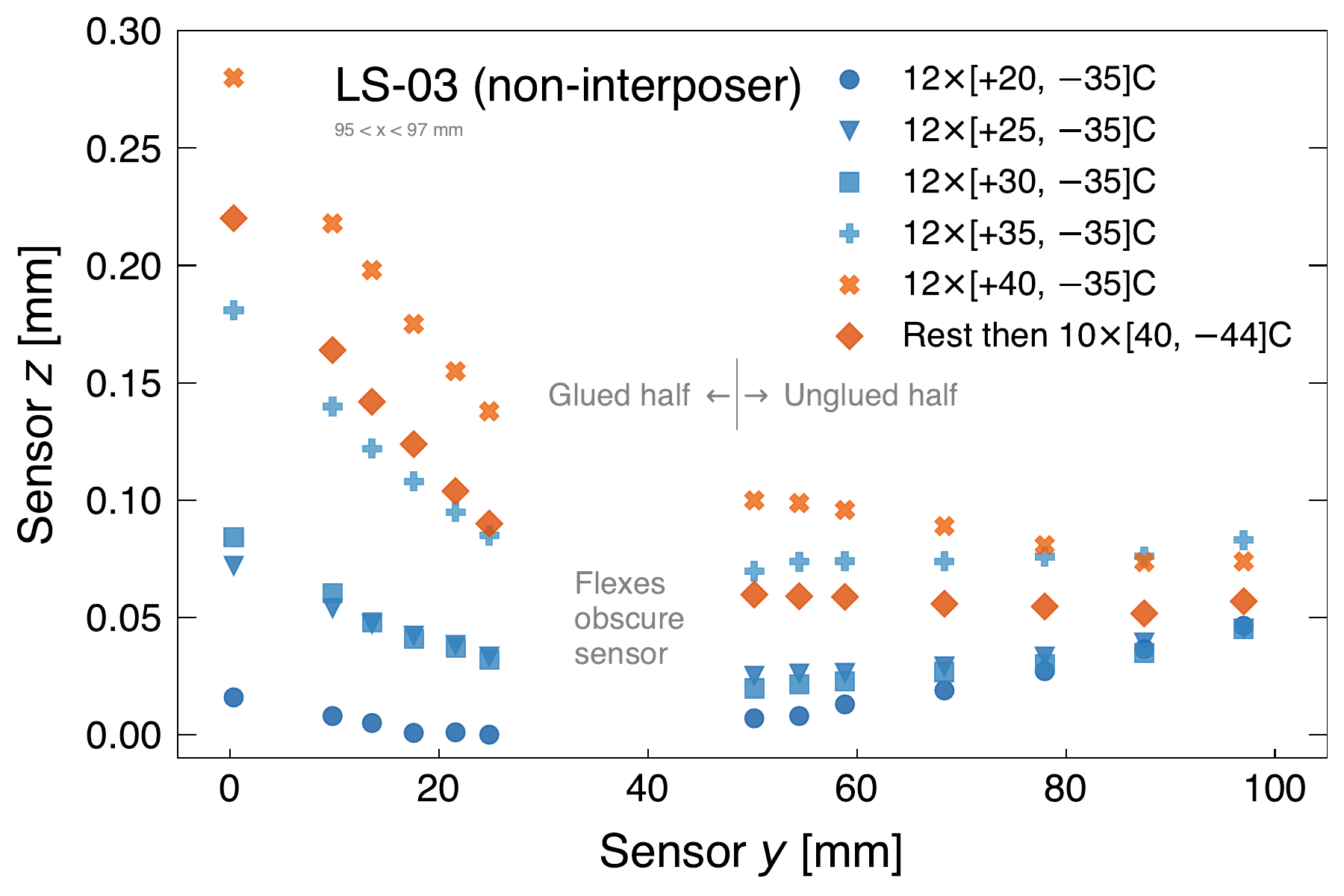}
    \caption{\label{fig:ls-03-bow-1dslice}LS-03 (no interposer)}
    \end{subfigure}%
    \begin{subfigure}{0.49\textwidth}
    \centering
    \includegraphics[width=\textwidth]{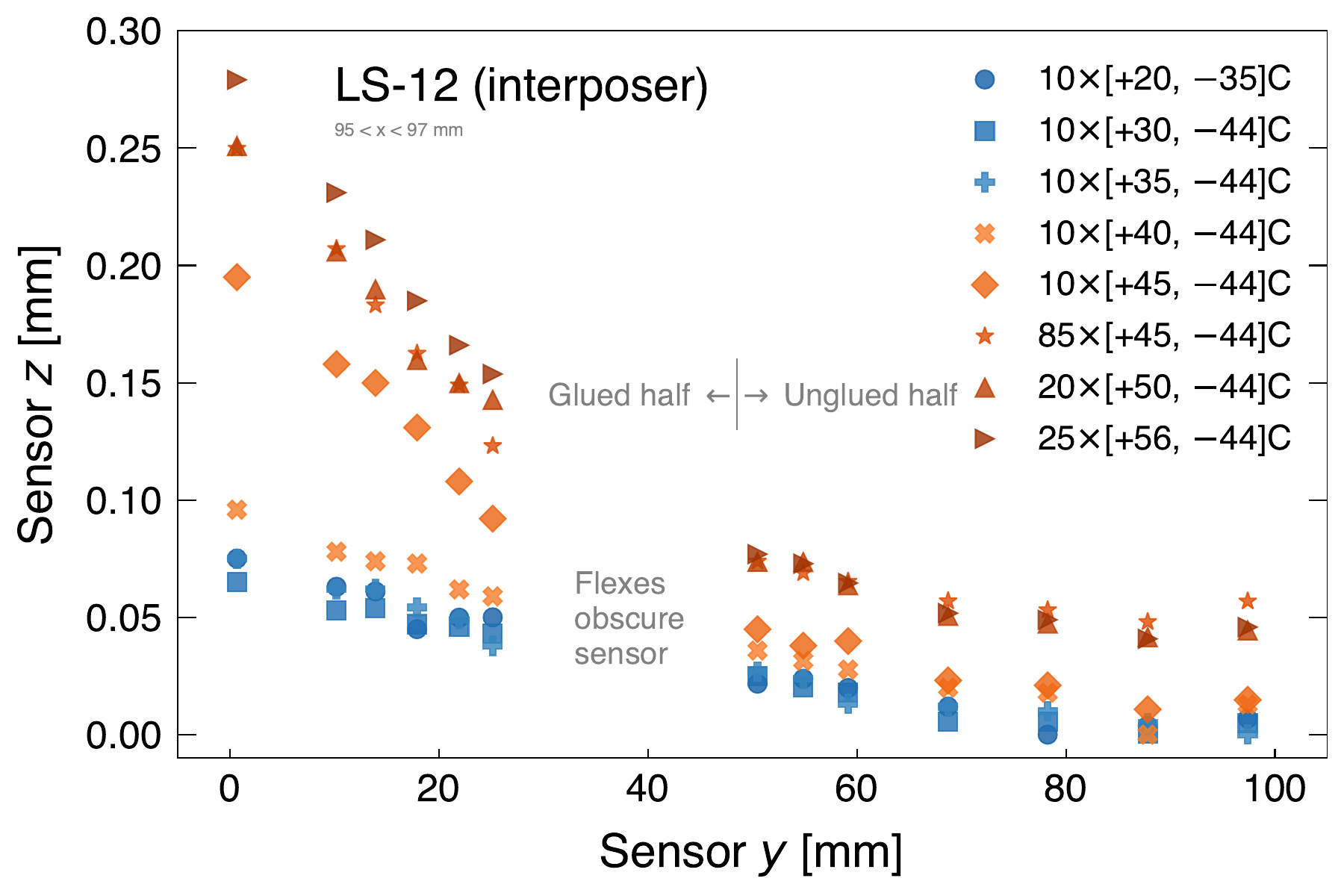}
    \caption{\label{fig:ls-12-bow-1dslice}LS-12 (with interposer)}
    \end{subfigure}
    \caption{\label{fig:1d-slices-interposer-vs-noninterposer}Comparison of sensor shape during MTC for (a) non-interposer vs (b) interposer module.
    This shows the $x\in [95,97]$~mm slice (side near HCCStar). 
    Flexes are glued for $y \in [25,50]$~mm, precluding sensor metrology. 
    The $y<50$~mm ($y>50$~mm) region is called the ``glued half'' (``unglued half''). 
    Metrology points after MTC with $T_\text{max} \geq 40^\circ$C are orange, else they are blue.
    }
\end{figure}

We can moreover illuminate the \emph{localised regions} of reduced deformation. 
Figure~\ref{fig:1d-slices-interposer-vs-noninterposer} compares the one-dimensional projection of sensor bow along the strip direction for a non-interposer (LS-03) and interposer module (LS-12).
In both cases, the sensor curvature is higher in the glued half ($y<50$~mm) than unglued half ($y>50$~mm) but remains within 100~$\mu$m for $T_\text{max} \leq 30^\circ$C. 
After MTC with $T_\text{max} = +35^\circ$C and $+40^\circ$C (blue and orange crosses), the LS-03 sensor curvature rises significantly for the glued half with little change in the unglued half (figure~\ref{fig:ls-03-bow-1dslice}).
Two-dimensional metrology 
shows the curvature is greatest at the sensor corners. 
The sensor corners elevate more when cooled, where suction cups in the handling frame exert downward pressure to cause a large lever arm about the sensor-epoxy-flex interface.
By contrast, the sensor metrology of the LS-12 interposer module is consistent after both $T_\text{max} = +35^\circ$C and $+40^\circ$C MTC (figure~\ref{fig:ls-12-bow-1dslice}).
The lack of sensor curvature change in the glued half of LS-12 further supports the interpretation that interposers mitigate thermal stress from the sensor-epoxy-flex CTE mismatch.

For the interposer modules, figure~\ref{fig:bowVsTemp} nonetheless shows a detectable bow increase $\Delta z\approx +30~\mu$m  when raising $T_\text{max}$ to $+45^\circ$C.
This is higher than $T_\text{max}=+35^\circ$C when this occurs for non-interposer modules, indicating increased headroom before sensor deformation appears. 
Figure~\ref{fig:ls-12-bow-1dslice} shows this arises from increased sensor curvature on the glued half of LS-12 after $T_\text{max}\geq +45^\circ$C MTC. 
It also requires substantially more cycles $50\times[+45, -44]^\circ$C to increase sensor bow to $\approx 150~\mu$m.
We subject SS-05 and LS-12 to extreme MTC: $95\times[+45, -44]^\circ$C before raising $T_\text{max}$ for  $20\times[+50, -44]^\circ$C and $25\times[+56, -44]^\circ$C.
The LS-12 bow reaches at most $174~\mu$m after $25\times[+56, -44]^\circ$C, which is lower than the non-interposer maxima where most reach $\approx 200~\mu$m after $[+40, -35]^\circ$C cycling. 
Interposer modules therefore require not only higher $T_\text{max}$ but also greater $N_\text{cycles}$ to reach over $150~\mu$m bow compared with $2\times[+35, -35]^\circ$C for non-interposer modules.
This delayed onset of sensor deformation further supports the promise of interposers as a stress mitigation strategy.

These two interposer modules undergo 200 thermocycles up to a $100^\circ$C temperature range between $[+56, -44]^\circ$C before terminating the study.
In the final few cycles of the $25\times [+56, -44]^\circ$C run, LS-12 shows HV breakdown around $320$~V during cold DAQ testing,  
which was absent in initial testing. 
This is a candidate electrical signature of thermal stress, but its delayed onset requires more extreme MTC conditions than pre-production modules.
No mechanical failures after such extreme thermocyling are observed by visual inspection. 

\section{\label{sec:summary}Conclusions}

In summary, this paper investigated mechanical problems observed during module thermocycling (MTC). 
We first alleviated non-thermal stress by levelling the MTC thermal chucks with shims. 
We also installed custom interlocks to enable long-term unattended thermocycling of five pre-production modules. 
Monitoring sensor bow while widening MTC temperature ranges, all modules exhibit significant bow increases with a mean of $146\pm 27~\mu$m after $[+40, -35]^\circ$C cycling compared with $[+20, -35]^\circ$C.
This indicates substantial thermal stress. 
Four modules saw sensor fractures after exposure to $[+40, -44]^\circ$C, while the remaining module survived 181 cycles up to $[+45, -44]^\circ$C. 

We then introduced stress-mitigating silicone and Kapton interposers to three modules via ad hoc assembly.
Detailed QC results established engineering feasibility and mitigation of the cold noise problem. 
After $[+40, -44]^\circ$C MTC, we observed an insignificant bow change $1\pm 10~\mu$m relative to $[+20, -44]^\circ$C. 
Two such modules underwent 200 cycles between $[+56, -44]^\circ$C with detectable bow increase but no fracturing.
Reduced sensor deformation contrasts with bow increases for \emph{every} non-interposer module and all but one fracturing.
Sensor shape analysis shows reduced curvature change by the flexes after MTC.
Together, these findings are interpreted as evidence that interposers are a promising strategy to mitigate thermal stress for module production.

\acknowledgments

We are very grateful for the many helpful and thoughtful discussions with various members of the ITk strip module collaboration.
We especially thank Will Fawcett for earlier collaboration on module pre-production, and Richard Shaw for engineering support in constructing the high-voltage supply and thermal interlock. 
The ATLAS Inner Tracker Strip Phase II upgrade at the Cavendish Laboratory in Cambridge is supported by the Science and Technology Facilities Council (STFC) of UK Research and Innovation (UKRI). 
The work of JL is generously supported by a Junior Research Fellowship at Trinity College, University of Cambridge.
Copyright 2025 CERN for the benefit of the ATLAS ITk Collaboration. CC-BY-4.0 license.

\bibliographystyle{JHEP}
\addcontentsline{toc}{section}{References}

\providecommand{\href}[2]{#2}\begingroup\raggedright\endgroup

\end{document}